\documentclass[onecolumn,secnumarabic,nofootinbib,superscriptaddress,aps]{revtex4-2}
\usepackage{color, xcolor, colortbl}
\usepackage{graphicx}
\usepackage{geometry}
\usepackage{amsthm,amsmath,amssymb,amsfonts}
\usepackage{dcolumn}
\usepackage{url}
\usepackage{epstopdf}
\usepackage{enumitem}
\usepackage{algorithm}
\usepackage{algpseudocode}
\usepackage{bm}
\usepackage{physics}
\usepackage[caption=false]{subfig}
\usepackage{appendix}
\usepackage{multirow}
\usepackage{braket}
\usepackage[english]{babel}
\usepackage{hyperref}
\usepackage[capitalize]{cleveref}
\usepackage{xpatch}
\usepackage{tikz}
\usepackage{adjustbox}
\usepackage{xspace}
\usetikzlibrary{quantikz}
\newcommand{\bvec}[1]{\mathbf{#1}}

\newcommand{\vv}{\bvec{v}}

\renewcommand{\Tr}{\operatorname{Tr}}

\renewcommand{\norm}[1]{\left\lVert#1\right\rVert}
\newcommand{\tnorm}[1]{{\left\vert\kern-0.25ex\left\vert\kern-0.25ex\left\vert#1\right\vert\kern-0.25ex\right\vert\kern-0.25ex\right\vert}}

\newcommand{\rd}{\,\mathrm{d}}
\newcommand{\Or}{\mathcal{O}}

\newtheorem{thm}{\protect\theoremname}
\newtheorem{lem}[thm]{\protect\lemmaname}

\providecommand{\definitionname}{Definition}
\providecommand{\assumptionname}{Assumption}
\providecommand{\corollaryname}{Corollary}
\providecommand{\lemmaname}{Lemma}
\providecommand{\propositionname}{Proposition}
\providecommand{\remarkname}{Remark}
\providecommand{\theoremname}{Theorem}
\providecommand{\problemname}{Problem}
\usepackage{bbm}

\makeatletter
\newenvironment{breakablealgorithm}
  {
   \begin{center}
     \refstepcounter{algorithm}
     \hrule height.8pt depth0pt \kern2pt
     \renewcommand{\caption}[2][\relax]{
       {\raggedright\textbf{\fname@algorithm~\thealgorithm} ##2\par}%
       \ifx\relax##1\relax 
         \addcontentsline{loa}{algorithm}{\protect\numberline{\thealgorithm}##2}%
       \else 
         \addcontentsline{loa}{algorithm}{\protect\numberline{\thealgorithm}##1}%
       \fi
       \kern2pt\hrule\kern2pt
     }
  }{
     \kern2pt\hrule\relax
   \end{center}
  }
\makeatother

\usetikzlibrary{fit}
\tikzset{%
  highlight/.style={rectangle,rounded corners,fill=blue!15,draw,fill opacity=0.3,thick,inner sep=0pt}
}

\usepackage{comment}

\newcommand{\cl}{\mathcal{L}}
\newcommand{\co}{\mathcal{O}}

\newcommand{\dt}{\Delta t}
\newcommand{\dw}{\Delta W}
\newcommand{\dz}{\Delta Z}

\newcommand{\ud}{\mathrm{d}}

\usepackage{xcolor,bm,physics}
\newcommand{\LL}[1]{\textcolor{magenta}{[LL:#1 ]}}

\newcommand{\re}[1]{{\color{black}#1}}
\graphicspath{{figures/}}

\begin{document}

\title{Simulating Open Quantum Systems Using Hamiltonian Simulations}

\newcommand{\DeptMath}{Department of Mathematics, University of California, Berkeley, CA 94720, USA}
\newcommand{\PenState}{
Department of Mathematics, Pennsylvania State University, State College, PA 16802, USA}
\newcommand{\LBLMath}{Applied Mathematics and Computational Research Division, Lawrence Berkeley National Laboratory, Berkeley, CA 94720, USA}
\newcommand{\CIQC}{Challenge Institute of Quantum Computation, University of California, Berkeley, CA 94720, USA}

\author{Zhiyan Ding} 
\email{zding.m@math.berkeley.edu}
\affiliation{\DeptMath}
\author{Xiantao Li}
\email{xxl12@psu.edu}
\affiliation{\PenState}
\author{Lin Lin}
\email{linlin@math.berkeley.edu}
\affiliation{\DeptMath}
\affiliation{\LBLMath}
\affiliation{\CIQC}

\begin{abstract}
We present a novel method to simulate the Lindblad equation, drawing on the relationship between Lindblad dynamics, stochastic differential equations, and Hamiltonian simulations. We derive a sequence of unitary dynamics in an enlarged Hilbert space that can approximate the Lindblad dynamics up to an arbitrarily high order. This unitary representation can then be simulated using a quantum circuit that involves only Hamiltonian simulation and tracing out the ancilla qubits. There is no need for additional postselection in measurement outcomes, ensuring a success probability of one at each stage. Our method can be directly generalized to the time-dependent setting. We provide numerical examples that simulate both time-independent and time-dependent Lindbladian dynamics with accuracy up to the third order.

\end{abstract}

\maketitle

\section{Introduction}

The Lindblad quantum master equation 
is a fundamental tool in studying open quantum systems~\cite{lindblad1976generators,gorini1976completely}. Unlike the time-dependent Schr\"odinger equation,  the Lindblad equation accounts for the effects of an environment on a quantum system by incorporating non-Hermitian operators that depict dissipative processes and jump operators that characterize environment noise.
Beyond its seminal applications in quantum electron dynamics \cite{Walls1994quantum,cohen1998atom,gardiner2000quantum,breuer2002theory}, 
the Lindblad equation, due to its universal representation property, has found extensive utility in various disciplines,  ranging from material science \cite{harbola2006quantum,di2007stochastic} to cosmology \cite{kiefer2007pointer}. 
Lindblad dynamics can also be used to describe circuit noise in quantum computing~\cite{pellizzari1995decoherence} and it underpins many quantum error-mitigation (QEM) strategies \cite{temme2017error,kandala2019error,endo2018practical,rossini2023single}. Recent advances have also leveraged Lindblad dynamics as an algorithmic tool for thermalizing quantum systems \cite{rall2022thermal,chifang2023quantum}, and for preparing ground states \cite{ding2023single}. 

As the range of applications for the Lindblad dynamics continues to expand, it becomes increasingly important to develop efficient and robust simulation methodologies. Classical simulation algorithms  \cite{breuer2002theory,biele2012stochastic,li2020exponential,CL21} are often hindered by a complexity that scales polynomially with the Hilbert-space dimension, resulting in exponential cost relative to the system size (such as the number of spins or qubits). In this context, quantum algorithms have emerged as promising alternatives that may reduce the cost exponentially. However, many of the current algorithms~\cite{KBG11,CL17,CW17,LW22,schlimgen2022quantum,chifang2023quantum}, particularly when high-order accuracy is required, can require many ancilla qubits, complicated quantum control logic for clock registers, and an involved amplitude-amplification procedure. These algorithms are thus much more intricate to implement compared to those designed for Hamiltonian simulation~\cite{berry2007efficient,berry2014exponential,Low2019hamiltonian,GilyenSuLowEtAl2019}.

\begin{figure}
    \centering
    \includegraphics[width=1\textwidth]{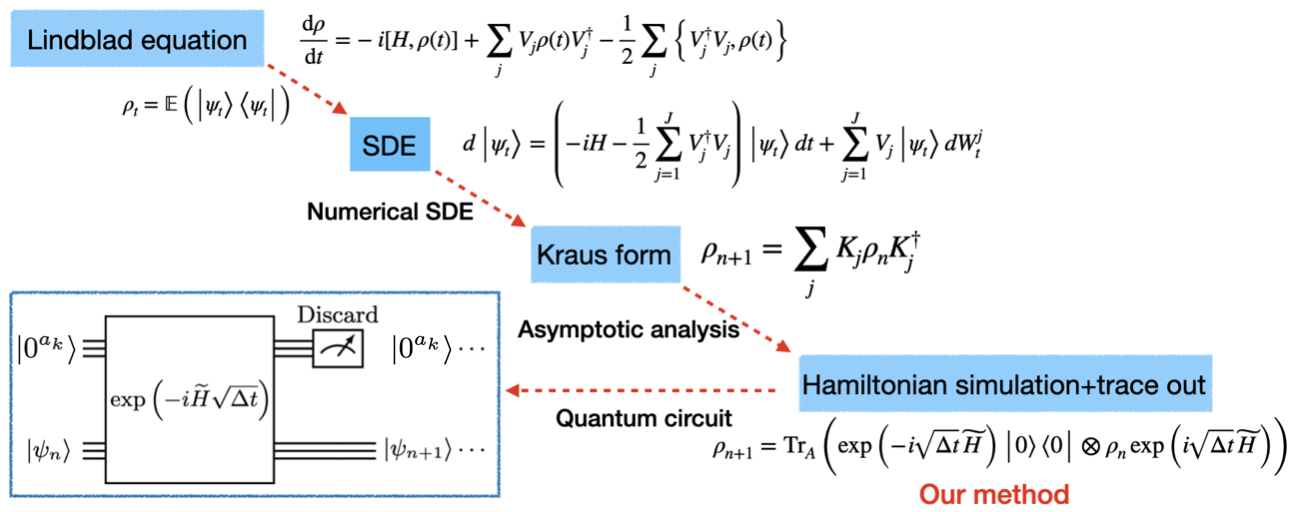}
    \caption{A flowchart illustrating the derivation of our numerical scheme and the quantum circuit (one step) for simulating the time-independent Lindbladian dynamics using the following steps: (1) unravelling the Lindblad equation into stochastic differential equations (SDEs). (2) express classical numerical SDE schemes as the Kraus-representation form for the density operator. (3) mapping the Kraus form to the dilated Hamiltonian in the Stinespring form. The simulation on the circuit advances a Hamiltonian simulation for a time duration of $\sqrt{\Delta t}$, after which the ancilla qubits are measured. The outcomes of these measurements on the ancilla qubit are disregarded and the ancilla qubits are subsequently reset to the state $\ket{0^{a_k}}$ in preparation for the next iteration. The inherent unitary and trace-out design ensures that the algorithm achieves a success probability of one, eliminating the need for any additional amplitude-amplification steps.}
    \label{fig:circuit}
\end{figure}

This paper presents a novel approach to simulating the Lindblad equation. Our method leverages the intimate relationship between Lindblad dynamics, stochastic differential equations (SDEs), and Hamiltonian simulations.
We show that, by adding extra ancilla qubits, the Lindblad dynamics can be incorporated into a unitary dynamics in a larger Hilbert space. Moreover, the unitary dynamics can be simulated using a quantum circuit that only involves Hamiltonian simulation and tracing out the ancilla qubits (see \cref{fig:circuit}). In this work, we present a systematic approach for constructing this unitary map and the corresponding Hamiltonian. Compared to other Lindblad simulation methods~\cite{wang2013solovay,CW17,CL17}, our proposed method has several distinct features: 

\begin{enumerate}
    \item Our numerical scheme reduces the Lindblad simulation problem to Hamiltonian simulations, for which many algorithms are available.
    \item When a unitary dynamics is constructed for the Hamiltonian simulation (e.g., via Trotterization), there is no need for additional postselection in measurement outcomes. The unitary evolution and the trace-out procedure guarantee that the success probability at each step is one, eliminating the need for amplitude-amplification procedures. 

    \item The algorithm can be systematically improved to achieve high-order accuracy.
    \item The algorithm can be easily generalized to time-dependent Lindbladians  in applications such as driven open quantum systems. Such direct generalization is highly nontrivial for many existing algorithms. 
\end{enumerate} 

Our procedure involves the following three steps, summarized in \cref{fig:circuit}. For simplicity, the Lindbladian dynamics are assumed to be time independent. The detailed explanation of the flowchart can be found in \cref{sec:main_thm_algorithm}. 
 
\begin{enumerate}

\item We unravel the Lindblad dynamics and reformulate them as SDEs. 

\item We use classical numerical SDE schemes and approximate the unraveled equation with an It\^o-Taylor expansion of an  arbitrary order of accuracy. This induces a Kraus representation of the dynamics of the density operator, which is completely positive.

\item Finally, instead of using the quantum algorithm due to \cite{LW22} to implement the Kraus form, we propose a new procedure that converts the Kraus form to the Stinespring form, detailing the construction of the Hamiltonian operator from the Kraus operators. This gives rise to a numerical scheme represented as a unitary dynamics that can be simulated through Hamiltonian simulation and trace-out. The resulting map is completely positive and trace preserving (CPTP).

\end{enumerate}

\subsection{Related works}
Wang et al. \cite{wang2013solovay} demonstrated how a single-qubit completely positive trace-preserving quantum channel can be approximated by simple quantum channels that can be simulated using only  one ancillary qubit. 
Kliesch et al.~\cite{KBG11} introduced the first quantum algorithm for simulating general Markovian open quantum systems. This algorithm has a complexity scaling of $\Or(t^2/\epsilon)$, where $t$ denotes the evolution time and $\epsilon$ represents the desired precision. The computational cost has been improved considerably in more recent works
\cite{CL17,CW17,LW22,kastoryano2023quantum}. In particular, the complexity of the algorithms in \cite{CW17,LW22,kastoryano2023quantum} is $\mathcal{O}\big(t \text{polylog} (t/\epsilon)\big)$, with a linear dependence on $t$ and polylogarithmic dependence on $\epsilon$. 
To our knowledge, all of the works focus on time-independent Lindbladian dynamics. In \cite{LW22}, the authors have suggested an extension of their method to time-dependent Lindblad dynamics, which emerges from rotating-wave approximations \cite{baker2018adaptive}. However, such an extension has not been fully explored, e.g.,  how to block encode the time-dependent Hamiltonians and jump operators.  
Schlimgen  et al. \cite{schlimgen2021quantum} have proposed to decompose Kraus operators into unitary operators that can be approximated by matrix exponentials. This approach has later been applied to the vectorized form of the Lindblad equation
\cite{schlimgen2022quantum}. The overall complexity, however, has not been presented. Andersson et al \cite{andersson2007finding} have explored how to construct the Kraus form for the quantum channel induced by the Lindblad dynamics, but without a full characterization of the numerical or model error. More importantly, this approach requires the input of the density matrix as a $d^2$-dimensional vector, with $d$ being the Hilbert-space dimension. Maintaining quantum speed-up with such a classical input is highly nontrivial. 
More recently, Patel and Wilde \cite{patel2023wave1,patel2023wave2} have proposed to encode the jump operators into a pure state $\ket{\psi}$, called a program state. Their algorithm is implemented through a quantum channel that involves both $\rho$ and $\psi$, followed by a trace-out step. For multiple jump operators, their approach follows a Trotter-type splitting \cite{CL17}, which is at most second order.  
 The work of Nakazato \cite{nakazato2006solution} has also studied the Kraus form, but with a focus on specific open quantum system models. Very recently, ~\cite{pocrnic2023quantum} has proposed a novel approach using repeated-interaction (RI) maps for approximating Lindblad dynamics. The simulation based on RI maps offers a first-order accuracy scheme for Lindblad simulations.

In the domain of Hamiltonian simulations, various algorithms with near-optimal query complexities suitable for different settings have been introduced. For instance, quantum signal processing~\cite{LowChuang2017} can reach the optimal query complexity for time-independent problems. In contrast, the truncated Dyson series~\cite{LowWiebe2019} is applicable to general time-dependent Hamiltonian simulation, but it is based on block encodings with complex control-logic operations. 
Although Trotterization does not achieve the optimal query complexity, it is more accessible in terms of its implementation (especially its lower-order versions). Taking this perspective into account, this work diverges notably from existing methods for simulating open quantum systems~\cite{CW17,LW22}, which are based on block encodings and entail complex control-logic operations. The nature and complexity of our algorithm for simulating open quantum systems resemble those of higher-order Trotter schemes used in Hamiltonian simulation. Combining our approach with different Hamiltonian simulation frameworks could lead to the development of new efficient Lindblad simulation algorithms.

\subsection{Organization}


The organization of the rest of the paper is as follows. In Section \ref{sec:preliminary}, we introduce essential notations, the relation between the Lindblad equation and the SDEs, along with classical numerical methods for solving SDEs. The main idea with the development of a first-order scheme is illustrated in Section \ref{sec:main_idea}. Our main results and quantum algorithms for simulating the Lindblad equation \eqref{eq:lindblad} are detailed in Section \ref{sec:main_thm_algorithm}. The performance of our algorithm is validated through various numerical experiments  in Section \ref{sec:nume}, for both time-independent and time-dependent Lindbladians.

Moreover, in Appendix \ref{sec:second_order_scheme}, we provide a detailed derivation of the time-independent second-order scheme, serving as a constructive example for our main results. For practical implementation, we provide formulations of the first-, second-, and third-order schemes (in both time-independent and time-dependent frameworks) in Appendix \ref{sec:all_scheme}. The technical proofs supporting our main results are found in Appendices \ref{sec:wt_R_alpha} and \ref{sec:H_existence}.



\section{Preliminaries}\label{sec:preliminary}

This paper uses capital letters for matrices and a curly font for superoperators. In particular, the identity map (superoperator) is denoted by $\mathcal{I}$ and the density operator (matrix), which is a positive-semidefinite (PSD) matrix with $\mathrm{Tr}(\rho)=1$, is represented by $\rho$. The vector or matrix 2-norm is denoted by $\|\cdot\| $: when $\vv$ is a vector, its 2-norm is denoted by $\|\vv\|$ and when $A$ is a matrix, its 2-norm (or operator norm) is denoted by $\| A\| $.

The trace norm (or Schatten $1$-norm) of a matrix $A$ is $\norm{A}_1=\Tr\left[\sqrt{A^{\dag}A}\right]$. Given a superoperator $\mathcal{M}$ that acts on operators (matrices in this paper), the induced $1$-norm is
\begin{equation}
\|\mathcal{M}\|_1=:\sup_{\|\rho\|_1\leq 1}\|\mathcal{M}(\rho)\|_1\,.
\end{equation}

The  main emphasis of the paper is on the approximation of  the Lindblad master equation \cite{lindblad1976generators,gorini1976completely},
\begin{equation}\label{eq:lindblad}
    \frac{d}{dt} \rho = \underbrace{-i[H, 
    \rho]}_{\mathcal{L}_H(\rho)} + \underbrace{\sum_{j=1}^J \left(V_j \rho V_j^\dagger - \frac{1}{2} \big\{V_j^\dagger V_j, \rho \big\} \right)}_{\mathcal{L}_V(\rho)}=:\mathcal{L}(\rho).
\end{equation}
Here $H\in \mathbb{C}^{d\times d}$ is the system Hamiltonian, and $V_j\in \mathbb{C}^{d\times d}$ are known as the jump operators that come from the interactions with the environment. 

The Gorini-Kossakowski-Lindblad-Sudarshan (GKLS) theorem~\cite{lindblad1976generators,gorini1976completely} states that if $\mathcal{L}$ is a Lindbladian with the form given in \eqref{eq:lindblad}, then $\exp(\mathcal{L}t)$ is a quantum channel, which means that it is a CPTP map that transforms one density operator into another. As a quantum channel, it is contractive under the trace distance~\cite[Theorem 9.2]{NielsenChuang2000}: for any two density operators $\rho_1$ and $\rho_2$, and any $t>0$, it holds that
\begin{equation}\label{eq:lindblad_contractive}
\big\|\exp(\mathcal{L}t)\rho_1-\exp(\mathcal{L}t)\rho_2\big\|_1 \leq \|\rho_1-\rho_2\|_1.
\end{equation}

To approximate the dynamics up to a given time $T$, one can divide the time interval into $N$ steps, $N \in \mathbb{N}$, with step size $\dt=T/N$. Thus it suffices to construct an approximation, here denoted by $\mathcal{M}_{\dt}\rho$, for a small step, e.g.,
\begin{equation}
\norm{\exp(\mathcal{L}\dt)\rho - \mathcal{M}_{\dt}[\rho]}_1 \leq C_k \dt^{k+1}, 
\end{equation}
for any density operator $\rho$ and some $k \geq 1$ with a constant $C_k$. The global error can then be deduced due to the contractive property, given in \eqref{eq:lindblad_contractive}. Specifically, if $\mathcal{M}_\dt[\cdot]$ is a quantum channel, we have  
\begin{equation}
\begin{aligned}
&\|\exp(\mathcal{L}T)\rho-(\mathcal{M}_{\dt})^N\rho\|_1\\
\leq &\|\exp(\mathcal{L}\dt)(\exp(\mathcal{L}(T-\dt))\rho-(\mathcal{M}_{\dt})^{N-1}\rho)\|_1\\
&+\|(\exp(\mathcal{L}\dt)-\mathcal{M}_{\dt})(\mathcal{M}_{\dt})^{N-1}\rho)\|_1\\
\leq &\left\|\exp(\mathcal{L}(T-\dt))\rho-(\mathcal{M}_{\dt})^{N-1}\rho\right\|_1+C_k\dt^{k+1}\\
\cdots & \cdots \\
\leq &C_k T\dt^{k},
\end{aligned}
\end{equation}
where we have repeated the method $N$ times to arrive at the last inequality. This gives us a $k$ th-order convergence, and we note that the final constant $C_k$ is independent of $T$.

\subsection{Unraveling the Lindblad equation using SDEs}
The solution to the Lindblad equation can be expressed through an SDE, which in turn also offers an intuitive description of a quantum dynamics subject to environmental noise.  Such a procedure is known as unraveling \cite{breuer2002theory} and, for this purpose,
we consider the stochastic Schr\"odinger equation, 
\begin{equation}\label{eqn:SDE_for_Lindblad}
    \ud \ket{\psi_t}=\left(-iH-\frac{1}{2}\sum^J_{j=1}V^\dagger_jV_j\right)\ket{\psi_t}\ud t+\sum^J_{j=1}V_j\ket{\psi_t}\rd W^j_t\,,
\end{equation}
where $\{W^j_t\}^J_{j=1}$ are independent Wiener processes, and the solutions are interpreted in It\^o's sense \cite{kloeden1992numerical}. 

The connection to the Lindblad equation \eqref{eq:lindblad} can be made by using It\^o's formula for $\ketbra{\psi_t}$ and taking the expectation, which yields
\begin{equation}\label{eqn:Lindblad_form}
    \frac{\rd \mathbb{E}(\ketbra{\psi_t})}{\rd t}=-i[H,\mathbb{E}\left(\ketbra{\psi_t}\right)]+\sum^J_{j=1}V_j\mathbb{E}\left(\ketbra{\psi_t}\right) V^\dagger_j-\frac{1}{2}\left\{V^\dagger_jV_j,\mathbb{E}\left(\ketbra{\psi_t}\right)\right\}\,.
\end{equation}
If the initial condition is $\mathbb{E}(\ketbra{\psi_0})=\rho_0$, then equation \eqref{eqn:Lindblad_form} is equivalent to the Lindblad equation~\eqref{eq:lindblad} with $\rho_t=\mathbb{E}(\ketbra{\psi_t})$.


In the classical regime, the aforementioned relationship serves as the basis for a  stochastic algorithm designed to simulate the Lindblad solution \cite{mora2018numerical,li2020exponential}. More specifically, the approach involves the following steps. First, several initial states ${\ket{\psi_{0,i}}}^N_{i=1}$ are randomly sampled from the density operator $\rho_0$. Next,  numerical simulations of \eqref{eqn:SDE_for_Lindblad} are performed for each initial state, evolving them up to time $T$. Finally, by averaging the resulting set of density matrices ${\ketbra{\psi_{T,i}}}$, one obtains an approximation to the solution $\rho_T$. 

\subsection{Numerical schemes for SDE}

Having reformulated the Lindblad dynamics using SDEs as in \cref{eqn:SDE_for_Lindblad}, we can leverage a wide variety of numerical techniques available in the literature for solving SDEs. In this paper, we mainly rely on the techniques described in~\cite[Chapter 14]{kloeden1992numerical}. The simplest among these methods is the Euler-Maruyama scheme, which, for any time step $\dt>0$, is given by 
\begin{equation}\label{eqn:first_order_SDE}
    \ket{\psi_{n+1}}=\ket{\psi_n}+\left(-iH-\frac{1}{2}\sum^J_{j=1}V^\dagger_jV_j\right)\ket{\psi_n}\dt+\sum^J_{j=1}V_j\ket{\psi_n}\sqrt{\dt}W^j=:L_{1,\dt}(\ket{\psi_n})\,,
\end{equation}
where $\{W^j\}^J_{j=1}$ are independent Gaussian random variables with zero expectation and unit variance. $\dt$ is a discretization of $\rd t$ in \eqref{eqn:SDE_for_Lindblad} and $\sqrt{\dt}W^j$ is a discretization of $\rd W^j_t$. This scheme provides a first-order approximation to the solution in the weak sense. Specifically, for $N\in\mathbb{N}$ and $T=N\dt$, we have 
\begin{equation}
\left\|\mathbb{E}(\ketbra{\psi_N})-\mathbb{E}(\ketbra{\psi_T})\right\|_1=\mathcal{O}\left(T\dt\right)\,.
\end{equation}
where $\ket{\psi_T}$ is the solution of \eqref{eqn:SDE_for_Lindblad} and the constant is independent of $\dt$.

Like ordinary differential equations (ODEs), higher-order numerical schemes can be obtained through a high-order expansion of SDEs. Due to the presence of the Brownian-motion terms,  the It\^o-Taylor expansion needs to be employed. This leads to many more terms when compared to such expansions from  ODEs (see the higher-order schemes in \cref{sec:all_scheme}). 


\section{Illustrative Demonstration Using a First-Order Algorithm}\label{sec:main_idea}

While numerical simulations of SDEs have been extensively explored in the literature, adapting these schemes directly for execution on a quantum computer presents challenges. For instance, the transformation from $\ket{\psi_n}$ to $\ket{\psi_{n+1}}$ in \eqref{eqn:first_order_SDE} is generally nonunitary, and there is no guarantee that $\ket{\psi_{n+1}}$ will remain a unit vector. On the other hand, since our objective is to simulate the Lindblad equation, it is not necessary to simulate every individual SDE trajectory~\eqref{eqn:SDE_for_Lindblad}. Instead, due to \eqref{eqn:SDE_for_Lindblad}, it suffices to simulate the ``expectation form'' of SDE~\eqref{eqn:SDE_for_Lindblad}.

We illustrate our main concept by deriving a first-order Lindblad simulation scheme from the Euler-Maruyama scheme~\eqref{eqn:first_order_SDE}. For simplicity, we assume $J=1$, i.e., there is only one jump operator. Using \eqref{eqn:first_order_SDE} and the property that $\mathbb{E}(W)=0$ and $\mathbb{E}(W^2)=1$, we obtain
\begin{equation}
\begin{aligned}
&\mathbb{E}(\ketbra{\psi_{n+1}})=\mathbb{E}\left(L_{1,\dt}[\ket{\psi_n}]\left(L_{1,\dt}[\ket{\psi_n}]\right)^\dagger\right)\\
=&\left(I+\left(-iH-\frac{1}{2}V^\dagger V\right)\dt\right)\mathbb{E}(\ketbra{\psi_{n}})\left(I+\left(iH-\frac{1}{2}V^\dagger V\right)\dt\right)\\
&+V\mathbb{E}(\ketbra{\psi_{n}})V^\dagger\dt\,.
\end{aligned}
\end{equation}
Let $F_0=I+\left(-iH-\frac{1}{2}V^\dagger V\right)\dt$, $F_1=V\sqrt{\dt}$, and $\rho_n=\mathbb{E}(\ketbra{\psi_{n}})$. The evolution from $\rho_n$ to $\rho_{n+1}$ is then expressed in the Kraus form:
\begin{equation}\label{eqn:Kraus_form}
\rho_{n+1}=:\mathbb{E}(\ketbra{\psi_{n+1}})=\mathcal{K}[\rho_n]=F_0 \rho_n F_0^\dagger+F_1 \rho_n F_1^\dagger\,.
\end{equation}
Furthermore, one also observes that
\begin{equation}
\rho_{n+1}=\rho_n+\mathcal{L}(\rho_n)\dt+\mathcal{O}(\dt^2)=\exp(\mathcal{L}t)\rho_n+\mathcal{O}(\dt^2)\,,
\end{equation}
where $\mathcal{L}$ is the Lindbladian that is defined in \eqref{eq:lindblad}. This equality implies that \eqref{eqn:Kraus_form} is a first-order scheme for the Lindblad equation. 

The above calculation shows that an SDE solver implies an approximation for the density matrix in the Kraus form.  
Next, to derive a first-order quantum simulation scheme, we further expand the Kraus form $\mathcal{K}$ in \cref{eqn:Kraus_form} into a Stinespring representation
\begin{equation}\label{eqn:S_representation_first}
\mathcal{K}[\rho]=\mathrm{Tr}_A\left(U\ketbra{0}\otimes \rho U^\dagger\right)=:\mathrm{Tr}_A\left(\begin{bmatrix}
        F_0 & \vdot \\
        F_1 & \vdot \\
    \end{bmatrix}\right.\ketbra{0}\otimes \rho \left.\begin{bmatrix}
        F_0 & \vdot \\
        F_1 & \vdot \\
    \end{bmatrix}^\dagger\right)\,.
\end{equation}
where $U$ is a unitary matrix that can be derived from Stinespring's factorization theorem. A key focus of this paper is on the construction of a Hamiltonian-generated unitary to approximate $U$, so that the algorithm can be implemented via a Hamiltonian simulation. In particular, we want to find a $2d\times 2d$ Hermitian matrix $\widetilde{H}$ such that
\begin{equation}\label{eqn:match_process}
\mathcal{K}[\rho]=\mathrm{Tr}_A\left(\exp(-i\sqrt{\dt}\widetilde{H})\ketbra{0}\otimes\rho\exp(i\sqrt{\dt}\widetilde{H})\right)+\mathcal{O}(\dt^2)\,,
\end{equation}
where the operator $\mathrm{Tr}_A$ traces out the ancilla qubit. We construct $\widetilde{H}$ that takes the following  form:
\begin{equation}
\widetilde{H}=\begin{bmatrix}
        H_0 & H^\dagger_1 \\
        H_1 & 0  \\
\end{bmatrix}\,,
\end{equation}
where $H_0$ is a Hermitian matrix. After applying Taylor expansion to $\exp(-i\sqrt{\dt}\widetilde{H})$ and matching $\co(1)$ and $\co(\dt)$ terms on both sides of \eqref{eqn:match_process}, we find that
\begin{equation}
H_0=\sqrt{\dt} H,\quad H_1=V\,.
\end{equation}
The above derivation suggests that the scheme
\begin{equation}\label{eqn:update_first_order}
    \rho_{n+1}=\mathrm{Tr}_A\left(\exp(-i\sqrt{\dt}\begin{bmatrix}
        \sqrt{\dt} H & V^\dagger \\
        V & 0  \\
\end{bmatrix})\ketbra{0}\otimes\rho_n\exp(i\sqrt{\dt}\begin{bmatrix}
        \sqrt{\dt} H & V^\dagger \\
        V & 0  \\
\end{bmatrix})\right)
\end{equation}
serves as a first-order approximation to the Lindblad equation \eqref{eq:lindblad}. This formula can be directly extended to the general case with multiple jump operators, simply by appending the additional jump operators along the first row and the first column.  Furthermore, the update process described in \eqref{eqn:update_first_order} only comprises a Hamiltonian simulation and a trace-out procedure, making it straightforward to implement and succeed with probability one. 

The above algorithm is similar to the first-order scheme in Ref.~\cite{CW17}, which uses first-order Trotter splitting to separate $\exp(\mathcal{L}_V \dt)$ and $\exp(\mathcal{L}_H \dt)$. Subsequently, it uses formulas analogous to those in \eqref{eqn:update_first_order} to simulate $\exp(\mathcal{L}_V \dt)$. 
However, it is difficult to extend the first-order scheme in \cite{CW17,pocrnic2023quantum} to high-order schemes. We note that the limitation of the first-order accuracy comes from two components: (i) the first-order approximation of the map $\exp(\mathcal{L}_V \dt)(\rho)$; and (ii) the first-order Trotter splitting used to separate $\exp(\mathcal{L}_V \dt)$ and $\exp(\mathcal{L}_H \dt)$. While the approximation of $\exp(\mathcal{L}_V \dt)(\rho)$ might be improved to a higher-order approximation, which is already not trivial, it is very difficult to avoid the first-order error caused by the first-order Trotter splitting. Unlike Hamiltonian simulation, the simulation of the dissipative part $e^{t\mathcal{L}_V}$ must have a non-negative $t$, meaning the simulation cannot go backward in time, since it does not constitute a CPTP map. However, for Trotter splitting beyond  second order with a real time variable $t$, a backward-in-time simulation is required \cite{BlanesCasas2005}. The method described in \cite{CW17} employs \eqref{eqn:update_first_order} merely as an illustrative example. The authors' primary algorithm is built upon the first-order method expressed in the Kraus form \eqref{eqn:Kraus_form} and the accuracy is boosted using a compression scheme. 
A key goal of this paper is to demonstrate that the Stinespring form, such as the one in \eqref{eqn:update_first_order}, paired with an appropriate dilated Hamiltonian, can be constructed to achieve arbitrary orders of accuracy.


\section{Main results}\label{sec:main_thm_algorithm}

In the previous section, the passage from  \cref{eqn:first_order_SDE}-\cref{eqn:Kraus_form} and \cref{eqn:S_representation_first} and then to \cref{eqn:update_first_order}, unveils a procedure to construct a Stinespring representation of the solution map with a Hamiltonian-generated unitary operator. Since numerical solutions for the SDE \eqref{eqn:SDE_for_Lindblad}
can be systematically constructed with an arbitrary order of accuracy, by taking expectations, we arrive at the Kraus-form approximation for simulating the Lindblad equation \eqref{eq:lindblad} to arbitrary order. Our main contribution is to extend the first-order scheme \eqref{eqn:update_first_order} to an arbitrarily high order. We present a family of  methods, as detailed in \eqref{eqn:main_scheme} and Algorithm \ref{alg:c_H}, to derive the unitary dynamics that approximate the Lindblad dynamics \eqref{eq:lindblad} to an arbitrarily high order. Moreover, the simulation of the unitary dynamics requires only Hamiltonian simulations and tracing out ancilla qubits, similar to \eqref{eqn:update_first_order}. 

Our main theoretical result is stated as follows.
\begin{thm}\label{thm:main}
Let $\left\|\mathcal{L}\right\|_{\mathrm{be}}=\left(1+\|H\|+\sum_j\|V_j\|^2\right)$. Given $k>0$, $\dt=\mathcal{O}(\left\|\mathcal{L}\right\|_{\mathrm{be}}^{-1})$, $N\in\mathbb{N}$, and $T=N\dt$. There exists a Hermitian matrix
\begin{equation}\label{eqn:H_widetilde_form}
\widetilde{H}=\ketbra{0}\otimes H_0+\sum^{S_k}_{j=1}\left(\ketbra{j}{0}H_j+\ketbra{0}{j}H^\dagger_j\right)\,,
\end{equation}
where the matrices $H_j \in \mathbb{C}^{d\times d}$, $H_0$ is Hermitian, the number of terms $S_k$ is upper  bounded by $(J+1)^{k+1}$, and $\norm{H_j}=\mathcal{O}\left(\left\|\mathcal{L}\right\|_{\mathrm{be}}\right)$.
 Furthermore, using $a_k\le \lceil (k+1)\log_2(J+1)\rceil$ ancilla qubits,
\begin{equation}\label{eqn:main_scheme}
\rho_{n+1}=\mathrm{Tr}_A\left(\exp(-i\sqrt{\dt}\widetilde{H})\ketbra{0^{a_k}}\otimes\rho_n\exp(i\sqrt{\dt}\widetilde{H})\right),
\end{equation}
is a $k$ th-order scheme for simulating the Lindblad equation~\eqref{eq:lindblad}, i.e.,
\begin{equation}
\left\|\rho_T-\rho_N\right\|=\mathcal{O}\left(T\left\|\mathcal{L}\right\|_{\mathrm{be}}^{k+1}\dt^{k}\right)\,,
\end{equation}
and the constant only depends on $k$ and $J$. 
\end{thm}
The proof of \cref{thm:main} is constructive. The Hermitian operator $\widetilde{H}$ in our construction will be called the \emph{dilated Hamiltonian}.
For any order $k>0$, we can always construct the corresponding Kraus-representation and Stinespring forms of the Lindblad dynamics~\eqref{eq:lindblad}. Specifically, we will propose a method to construct each block of $\widetilde{H}$ (denoted as $H_0, H_1, \ldots, H_{S_k}$) using a polynomial of $H$, $V_j$, $V_j^\dagger$, and $\Delta t^{1/2}$ with the maximum degree of $\mathrm{poly}(k)$. According to the above theorem, our algorithm requires $\mathcal{O}(k\log(J+1))$ ancilla qubits to generate a $k$ th-order scheme, which is slightly fewer than the $\Omega(k\log((J+1)k))$ ancilla qubits needed in \cite{LW22}.

In~\cite[Theorem 4]{CW17}, a lower bound on the total Hamiltonian simulation time is proved using the amplitude-damping process. It asserts that discretizing Lindblad dynamics into $N$ stages requires a minimum total Hamiltonian evolution time of $\Omega(\sqrt{N})$. In \cref{thm:main} with $N = 1/\Delta t$ (assuming the final time $T=1$ for simplicity), this implies that the required total Hamiltonian simulation time must be at least $\mathcal{O}(1/\sqrt{\Delta t})$. In \cref{eqn:main_scheme} the Hamiltonian simulation time step is $\sqrt{\Delta t}$, resulting in a total simulation time of $\sqrt{\Delta t}/\Delta t = 1/\sqrt{\Delta t}$, which agrees the aforementioned lower bound.

In practical applications, the implementation of $\exp(-i\widetilde{H}t)$ relies on the assumptions made about the oracles for $H$ and $V_j$. Assuming that $H$ and $V_j$ can be decomposed into a sum of local operators, we can then decompose each $H_j$ into a sum of local operators. This decomposition enables the implementation of $\exp(-i\widetilde{H}t)$ using e.g., a high-order Trotter formula. In this case, the complexity depends on how complicated $H$ and $V_j$ are. An alternative method of implementation involves utilizing block encoding. In \cref{sec:implement}, we explore a specific approach to implement the block encoding of $\widetilde{H}$ assuming the block encoding of $H$ and $H_j$. This provides a method for implementing $\exp(-i\widetilde{H}t)$ using block-encoding-based Hamiltonian simulation algorithms.   Alternative methods for efficiently implementing $\exp(-i\widetilde{H}t)$ are an important direction for future research.

\subsection{Overview of the main algorithm}\label{sec:main_algorithm}

In this section, we describe the construction of our main simulation algorithm, focusing on deriving the $k$ th-order scheme for the time-independent Lindblad equation. We outline the general procedure for constructing the Hamiltonian $\widetilde{H}$ for any $k$ and in Appendix \ref{sec:second_order_scheme} we provide a specific example of a second-order scheme for time-independent Lindbladian dynamics. In \cref{sec:all_scheme}, we extend our approach to time-dependent Lindblad equations and present the explicit forms of $\widetilde{H}$ for the first- 
 to third-order schemes, covering both time-dependent and time-independent scenarios. 
    
We first note that the simulation algorithm for~\eqref{eq:lindblad} is straightforward after obtaining $\widetilde{H}$ (see \cref{fig:circuit}). Given a required order $k>0$, after finding the Hamiltonian $\widetilde{H}$ such that
\begin{equation}
\exp(\mathcal{L}\dt)\rho=\mathrm{Tr}_A\left(\exp(-i\sqrt{\dt}\widetilde{H})\ketbra{0^{a_k}}\otimes\rho\exp(i\sqrt{\dt}\widetilde{H})\right)+\mathcal{O}(\dt^{k+1})\,,
\end{equation}
our numerical scheme is 
\begin{equation}
\rho_{n+1}=\mathrm{Tr}_A\left(\exp(-i\sqrt{\dt}\widetilde{H})\ketbra{0^{a_k}}\otimes\rho_n\exp(i\sqrt{\dt}\widetilde{H})\right)+\mathcal{O}(\dt^{k+1})\,.
\end{equation}
The trace-out process can be accomplished by measuring and resetting the ancilla qubit. 


Now, we focus on our approach to constructing the dilated Hamiltonian $\widetilde{H}$ in \cref{eqn:H_widetilde_form}. Similar to the derivation of the first-order scheme in \cref{sec:main_idea}, we follow three steps to generate a $k$ th-order scheme, 
\begin{enumerate}[wide, labelindent=0pt]
    \item[\bf Step 1.] Formulate the weak scheme of order $k$ for SDEs in \eqref{eqn:SDE_for_Lindblad}. Find a random linear operator $L_{k,\dt}:\mathbb{C}^d\rightarrow \mathbb{C}^d$ that generalizes \eqref{eqn:first_order_SDE}, such that for any unit vector $\ket{\psi}$, 
    \begin{equation}\label{eqn:k-th_order_condition_L}
    \left\|\mathbb{E}\left(L_{k,\dt}[\ket{\psi}]\left(L_{k,\dt}[\ket{\psi}]\right)^\dagger\right)-\mathbb{E}\left(\ketbra{\psi(\dt)}\right)\right\|_1=\mathcal{O}((\dt)^{k+1})\,,
    \\ \end{equation}
    where $\ket{\psi(\dt)}$  is a realization of the solution of \eqref{eqn:SDE_for_Lindblad} with $\ket{\psi(0)}=\ket{\psi}$. We recall that $\rho(\dt)=\mathbb{E}\left(\ketbra{\psi(\dt)}\right)$ is the solution of the Lindblad equation with $\rho(0)=\mathbb{E}\left(\ketbra{\psi(0)}\right)$.

    We note that there are many approaches to designing a $k$ th-order weak formulation for SDE \eqref{eqn:SDE_for_Lindblad}. In \cref{sec:pf_thm}, we will present the It\^o-Taylor-expansion approach  from~\cite[Chapter 14]{kloeden1992numerical}.
        
    \item[\bf Step 2.] Formulate the $k$ th-order Kraus form: From the operator $L_{k,\dt}$, find a sequence of Kraus operators  $\{F_j\}^{S_k}_{j=0}$, where $S_k\leq (J+1)^k$, such that
    \begin{equation}\label{eqn:k-th_order_condition_F}
    \mathbb{E}\left(L_{k,\dt}[\ket{\psi}]\left(L_{k,\dt}[\ket{\psi}]\right)^\dagger\right)=\sum^{S_k}_{j=0}F_j\ketbra{\psi}F^\dagger_j+\mathcal{O}\left((\dt)^{k+1}\right)\,.
    \end{equation}
    The above equation directly implies that the trace-preserving property holds approximately,
\begin{equation}\label{eqn:trace_preserving}
     \sum_{j=0}^{S_k} F_j^\dag  F_j = I +  \co(\dt^{k+1})\,. 
\end{equation}

    We can explore various methods to construct the Kraus form mentioned above. In \cref{sec:pf_thm}, we will discuss one approach to obtain the Kraus form associated with a $k$ th-order weak scheme for the SDEs.
    With the Kraus form ready, the algorithms in \cite{CW17,LW22} can be directly used to simulate the Lindblad dynamics by implementing the Kraus form. Therefore, the unraveling approach provides an alternative to obtaining a higher-order approximation expressed in Kraus form, without using Dyson series and numerical quadrature.  More importantly, here we take a different path forward, by converting 
  the Kraus form to a Stinespring form, thereby enabling simulations of the Lindblad dynamics through  Hamiltonian simulations.

\item[\bf Step 3.] Construct the dilated Hamiltonian $\widetilde{H}$. Find a sequence of matrices $\{H_j\}^{S_k}_{j=0}$ such that
\begin{equation}\label{eqn:F_H_match}
\sum^{S_k}_{j=0}F_j\ketbra{\psi}F^\dagger_j=\mathrm{Tr}_A \left( \exp(-i\sqrt{\dt}\widetilde{H}) \ketbra{0^{a_k}}\otimes \ketbra{\psi} \exp(i\sqrt{\dt}\widetilde{H}) \right)+\mathcal{O}((\dt)^{k+1})\,,
\end{equation}
where the Hermitian matrix $\widetilde{H}=\ketbra{0}H_0+\sum^{S_k}_{j=1}\ketbra{j}{0}H_j +\ketbra{0}{j}H^\dagger_j$. This is achieved through asymptotic analysis. This versatile approach is applicable not only when the Kraus form is derived from an SDE integrator but also in situations in which the Kraus form emerges from alternative derivations.
\end{enumerate}

\subsection{Proof of the main theorem: Construction of the dilated Hamiltonian \texorpdfstring{$\widetilde{H}$}{TEXT}}~\label{sec:pf_thm}

In this section, we detail the strategies to accomplish the preceding three steps, which provide a constructive proof of Theorem \ref{thm:main}. 
The algorithm to construct $\widetilde{H}$ is summarized in Algorithm \ref{alg:c_H}.  
\begin{breakablealgorithm}
      \caption{Construction of the dilated Hamiltonian $\widetilde{H}$}
  \label{alg:c_H}
  \begin{algorithmic}[1]
    \Statex \textbf{Input:} Desired order: $k$; Time step: $\dt$; Hamiltonian: $H$; Jump operators: $\{V_j\}$;
    \Statex \textbf{Output:} $\widetilde{H}$.
    \State Formulate a $k$ th-order SDE scheme following \cref{eqn:L}.    
   \State Produce  the corresponding $k$ th-order Kraus using \cref{eqn:Kraus}.

   \State Construct the dilated Hamiltonian $\widetilde{H}$ based on the pathway detailed in \cref{eqn:derivation_road} and \cref{fig:flowchart_match} in \cref{sec:H_existence}.

    \end{algorithmic}
\end{breakablealgorithm}

In the following part of the derivation, we simplify our notation by omitting the subindex of $L_{k,\dt}$ and denoting it as $L$. We also define 
\begin{equation}\label{eqn:V_0}
    V_0=-iH-\frac{1}{2}\sum^J_{j=1}V^\dagger_jV_j\,,
\end{equation}
which is responsible for the non-Hermitian part of the 
Lindblad dynamics. We will not include the subscript of $\ket{\psi_n}$ in the following proof for the sake of simplicity.

\textbf{Step 1: Formulate the weak scheme of order $k$ for the SDE \eqref{eqn:SDE_for_Lindblad}.} 

The $k$ th-order weak scheme has been thoroughly investigated in the classical numerical SDE literature. Here, we employ the scheme derived from the It\^o-Taylor expansion as presented in~\cite[Chapter 14]{kloeden1992numerical}. Toward this end, we define two sets of multi-indices
\begin{equation}
\Gamma_k=\{\alpha=(j_1,j_2,\cdots,j_{|\alpha|})\in\{0,1,2,\cdots,J\}^{\otimes |\alpha|}:|\alpha|\leq k\}\,,
\end{equation}
and
\begin{equation}
\Gamma_{k/0}=\Gamma_k\setminus \{\alpha=\{0\}^{\otimes |\alpha|}:|\alpha|\leq k\}\,,
\end{equation}
where $|\alpha|$ is the number of components of the multi-index $\alpha$. These indices are necessary to keep track of the different components of the Brownian motion $W_j(t).$
A scheme of weak order $k$ can be expressed using multiple integrals over $0\le s_1\le s_2\le \cdots\le s_k\le \Delta t$, 
\begin{equation}\label{eqn:L}
\begin{aligned}
    L[\ket{\psi}]=&\ket{\psi}+\sum_{\alpha\in\Gamma_k}\left(V_{j_1}V_{j_2}\cdots V_{j_{|\alpha|}}\ket{\psi}\right)\int^{\dt}_{0}\int^{s_{|\alpha|}}_{0}\int^{s_{{|\alpha|}-1}}_{0}\cdots \int^{s_2}_0 \ud W^{j_1}_{s_1}\ud W^{j_2}_{s_2}\cdots \ud W^{j_{|\alpha|}}_{s_{|\alpha|}}\\
    =&\sum^k_{j=0}\frac{(\dt)^j}{j!}V^j_0\ket{\psi}+\sum_{\alpha\in\Gamma_{k/0}}R_\alpha\textbf{V}_\alpha\ket{\psi}
\end{aligned}
\end{equation}
where we set $\ud W^0_s=\ud s$, $\textbf{V}_\alpha=V_{j_1}V_{j_2}\cdots V_{j_{|\alpha|}}$ denotes a product of the jump operators, and the sequence of random variables $\{R_\alpha\}_{\alpha\in\Gamma_{k/0}}$ corresponds to multiple It\^o stochastic integrals as follows:
\begin{equation}\label{eqn:R_alpha}
R_\alpha=\int^{\dt}_{0}\int^{s_{|\alpha|}}_{0}\int^{s_{{|\alpha|}-1}}_{0}\cdots \int^{s_2}_0 \ud W^{j_1}_{s_1}\ud W^{j_2}_{s_2}\cdots \ud W^{j_{|\alpha|}}_{s_{|\alpha|}}\,.
\end{equation}
According to~\cite[Theorems 14.5.1, 14.5.2]{kloeden1992numerical}\footnote{Strictly speaking,  \eqref{eqn:k-th_order_condition_L} is not a direct result of these two theorems but can be shown by the proof of Theorem 14.5.2.}, the direct expansion given in \eqref{eqn:L} induces a $k$ th-order weak scheme that satisfies the desired order condition given in \eqref{eqn:k-th_order_condition_L}. In addition, when $\dt=\mathcal{O}\left(\|\mathcal{L}\|_{\mathrm{be}}\right)$, we have
\begin{equation}
\left\|\mathbb{E}\left(L[\ket{\psi}]\left(L[\ket{\psi}]\right)^\dagger\right)-\mathbb{E}\left(\ketbra{\psi(\dt)}\right)\right\|_1=\mathcal{O}\left(\|\mathcal{L}\|_{\mathrm{be}}^k(\dt)^{k+1}\right).
\end{equation}

\textbf{Step 2: Formulate the $k$-th order Kraus form.}

In the second step, we construct the Kraus form of $k$ th-order from the It\^o-Taylor-expansion method in \eqref{eqn:L}. As a preparation, we introduce some notation and definitions for the terms with multi-indices. Note that the zero components in $\alpha$ indicate a standard integration over $t$, while nonzero components correspond to stochastic integrals.  Given $\alpha\in\Gamma_k$, let $\alpha^+$ be the multi-index obtained by removing all components of $\alpha$ that are equal to zero. For example, if $\alpha=(1,0,2,1)$, then we have 
\[
\alpha^+=(1,0,2,1)^+=(1,2,1)\,.
\]
We define $l_{=0}(\alpha)$ as the number of zero elements, which means that $l_{=0}(\alpha)=|\alpha|-|\alpha^+|$. According to~\cite[Chapter 5, Lemma 5.7.2]{kloeden1992numerical}, given $\alpha,\alpha'\in\Gamma_k$, we have
\begin{equation}\label{eqn:ploy_cov}
\mathbb{E}\left[R_{\alpha}R_{\alpha'}\right]=C_{\alpha,\alpha'}\Delta t^{|\alpha|+|\alpha'|-|\alpha^+|}\textbf{1}_{\alpha^+=(\alpha')^+},\quad C_{\alpha,\alpha'}= \mathcal{O}(1)\,.
\end{equation}
Here, $\textbf{1}_{\alpha^+=(\alpha')^+}$ stands for the indicator function and $C_{\alpha,\alpha'}$ is a factor that depends on the indices $\alpha$ and $\alpha'$ but not on $\dt$. In addition, $|C_{\alpha,\alpha'}|\leq 1$ for all $\alpha,\alpha'$. Based on \eqref{eqn:ploy_cov}, we define the normalization of $R_{\alpha}$ by the step size $\dt$:
\begin{equation}\label{Rna}
R_{\mathrm{n},\alpha}=R_{\alpha}\dt^{-\frac{|\alpha|+l_{=0}(\alpha)}{2}}\,.
\end{equation}
As a result of this rescaling, we can work with a set of Gaussian random variances $R_{\mathrm{n},\alpha}$ with mean zero and covariance independent of $\dt.$
In particular, we can  rewrite $L[\ket{\psi}]$ in \eqref{eqn:L} as
\[
L[\ket{\psi}]=\sum^k_{j=0}\frac{(\dt)^j}{j!}V^j_0\ket{\psi}+\sum_{\alpha\in\Gamma_{k/0}}R_{\mathrm{n},\alpha} \left(\dt^{\frac{|\alpha|+l_{=0}(\alpha)}{2}}\textbf{V}_\alpha\ket{\psi}\right)\,.
\]
Here $\mathbb{E}(R^2_{\mathrm{n},\alpha})=C_{\alpha,\alpha'}$.

Note that even though the expected value of $R_{\mathrm{n},\alpha}$ is zero, the expected value of $R_{\mathrm{n},\alpha} R_{\mathrm{n},\alpha'}$ may not be equal to zero; i.e., in general, these random variables are correlated. Specifically, 
\begin{equation}
\mathbb{E}(R_{\mathrm{n},\alpha} R_{\mathrm{n},\alpha'})\neq 0\,.
\end{equation}
Thus, if we naively define $K_\alpha=\sqrt{\mathbb{E}(R^2_{\mathrm{n},\alpha})}\dt^{\frac{|\alpha|+l_{=0}(\alpha)}{2}}\textbf{V}_\alpha$, we will encounter some cross terms in the expansion of the Kraus form, leading to a nondiagonal Kraus form. To overcome this difficulty, we introduce the following lemma to orthogonalize the noise term.
\begin{lem}\label{lem:R_alpha}
Let $R_{\mathrm{n},\alpha}$ be defined in \eqref{Rna}. There exists a sequence of random variables $\left\{\widetilde{R}_\alpha\right\}_{\alpha\in\Gamma_{k/0}}$ that satisfy the following conditions:
\begin{itemize}
    \item Each $R_{\mathrm{n},\alpha}$ is a linear combination of $\widetilde{R}_{\alpha'}$ such that 
    \begin{equation}\label{eqn:R_alpha_lem}
R_{\mathrm{n},\alpha}=\sum_{\alpha'\in\Gamma_{k/0}}c_{\alpha,\alpha'}\widetilde{R}_{\alpha'}\,,
    \end{equation}
    where $c_{\alpha,\alpha'}$ is a constant independent of $\dt$. In addition, $\sum_{\alpha'} |c_{\alpha,\alpha'}|^2=\mathbb{E}(R^2_{\mathrm{n},\alpha})$ and $c_{\alpha,\alpha'}=0$ if $\alpha^+\neq (\alpha')^+$. 
    \item For any $\alpha$, $\mathbb{E}\left(\widetilde{R}_\alpha\right)=0$. In addition, $\widetilde{R}_\alpha$ is either zero or $\mathbb{E}(\widetilde{R}^2_\alpha)=1$. 
    \item For any $\alpha\neq \alpha'\in\Gamma_k$, we have $\mathbb{E}\left(\widetilde{R}_\alpha \widetilde{R}_{\alpha'}\right)=0$, i.e., they are uncorrelated. 
\end{itemize}
\end{lem}

The proof of \cref{lem:R_alpha} is in Appendix \ref{sec:wt_R_alpha}. With this new expression for the noise terms, we can plug \cref{eqn:R_alpha_lem} from Lemma \ref{lem:R_alpha}  into \eqref{eqn:L} and obtain
\begin{equation}\label{eqn:Kraus_prep}
    L_{k,\Delta t}[\ket{\psi}]
    =\sum^k_{j=0}\frac{(\dt)^j}{j!}V^j_0\ket{\psi}+\sum_{\alpha\in\Gamma_{k/0}}\widetilde{R}_\alpha\left(\sum_{\alpha'\in\Gamma_k}c_{\alpha',\alpha}\dt^{\frac{|\alpha'|+l_{=0}(\alpha')}{2}}\textbf{V}_{\alpha'}\right)\ket{\psi}
\end{equation}

We are now in a position to derive a Kraus form. We define
\begin{equation}\label{eqn:F_formulation}
F_0=\sum^k_{j=0}\frac{(\dt)^j}{j!}V^j_0,\quad F_{\alpha}=\left(-i\sum_{\alpha'\in\Gamma_k}c_{\alpha',\alpha}\dt^{\frac{|\alpha'|+l_{=0}(\alpha')}{2}}\textbf{V}_{\alpha'}\right)\textbf{1}_{\widetilde{R}_\alpha\neq 0},\quad \forall \alpha\in \Gamma_{k/0}\,.
\end{equation}


In light of \eqref{eqn:Kraus_prep}, we obtain an approximation of the density-operator in a Kraus form,
\begin{equation}\label{eqn:Kraus}
\mathbb{E}\left(L[\ket{\psi}]\left(L[\ket{\psi}]\right)^\dagger\right)=F_0\ketbra{\psi}F^\dagger_0+\sum_{\alpha\in\Gamma_{k/0}}F_\alpha\ketbra{\psi}F^\dagger_\alpha\,,
\end{equation}
which satisfies \eqref{eqn:k-th_order_condition_F}. We note that the total number of Kraus operators is at most $\frac{(J+1)^{k+1}-1}{J}-k$.

\textbf{Step 3: Construct the dilated Hamiltonian $\widetilde{H}$.}

We start by ordering and expressing Kraus operators by the powers of $\dt$, i.e.,  in an asymptotic form:
\begin{equation}\label{asymp-kraus}
    \begin{aligned}
        F_0 & = I + \dt Y_{0,0} + \dt^2 Y_{0,1} +\dt^3 Y_{0,2}+\cdots+\dt^{k}Y_{0,k-1},  \\  
        F_j &=-i \left( {\dt}^{1/2} Y_{j,0} + {\dt}^{3/2} Y_{j,1} + {\dt}^{5/2} Y_{j,2} + \cdots+\dt^{k-1/2}Y_{j,k-1}\right),   \quad j=1,2,\cdots, s_k, \\
        F_j&= -i \left( \dt Y_{j,0} + \dt^2 Y_{j,1} + \cdots+\dt^{k-1}Y_{j,k-2}\right),  \quad j=s_k+1,\cdots, S_k.
    \end{aligned}
\end{equation}
Here, we separate those Kraus operators with integer powers of $\dt$ from those with half powers of $\dt$. We note that $S_k+1$ equals to the number of Kraus operators. Thus, $S_k\leq \frac{(J+1)^{k+1}-1}{J}-k-1< (J+1)^{k+1}$.

From \eqref{eqn:k-th_order_condition_L} and \eqref{eqn:k-th_order_condition_F}, we see that $\sum^{S_k}_{j=0}F_j\rho F^\dagger_j$ is a $k$ th-order approximation of a Lindblad equation and can be expanded into Stinespring form, meaning that 
\begin{equation}\label{eq:lindblad_approximatios_k}
\begin{aligned}
     e^{\cl\dt} \rho=&\sum^{S_k}_{j=0}F_j\rho F^\dagger_j + \co((\dt)^{k+1})=\mathrm{Tr}_A\left(U\ketbra{0^{a_k}}\otimes \rho U^\dagger\right)+ \co((\dt)^{k+1})\\
     =:&\mathrm{Tr}_A\left(\begin{bmatrix}
        F_0 & \vdot & \cdots & \vdot\\
        F_1 & \vdot & \cdots & \vdot\\
        \vdots  & \vdots& \ddots & \vdots\\
        F_{S_k} & \vdot & \cdots & \vdot\;
    \end{bmatrix}\right.\ketbra{0}\otimes \rho \left.\begin{bmatrix}
        F_0 & \vdot & \cdots & \vdot\\
        F_1 & \vdot & \cdots & \vdot\\
        \vdots  & \vdots& \ddots & \vdots\\
        F_{S_k} & \vdot & \cdots & \vdot\;
    \end{bmatrix}^\dagger\right)+ \co((\dt)^{k+1})\,.
\end{aligned} 
\end{equation}
where $U$ is a unitary matrix that can be constructed by Stinespring's factorization theorem. 

Now, we are ready to introduce the following lemma that implies the existence of the dilated Hamiltonian  $\widetilde{H}$.
\begin{lem}\label{lem:H_existence} Given the Kraus operators $\{F_j\}^{S_k}_{j=0}$ in \eqref{asymp-kraus}, there exists $\widetilde{H}$ such that 
\begin{equation}\label{eqn:H_lemma}
\sum^{S_k}_{j=0}F_j\ketbra{\psi}F^\dagger_j=\mathrm{Tr}_A \left( \exp(-i\sqrt{\dt}\widetilde{H}) \ketbra{0^{a_k}}\otimes \ketbra{\psi} \exp(i\sqrt{\dt}\widetilde{H}^\dagger) \right)+\mathcal{O}((\dt)^{k+1})\,,
\end{equation}
Furthermore, $\widetilde{H}$ can be written as \eqref{eqn:H_widetilde_form} with
\begin{equation}\label{asymp-ham}
    \begin{aligned}
        H_0 & = \dt^{1/2} X_{0,0} + \dt^{3/2} X_{0,1} \cdots + \dt^{k-1/2}X_{0,k-1}, \quad \\  
        H_j &= X_{j,0} + {\dt} X_{j,1}  + \cdots +\dt^{k-1}X_{j,k-1}, \quad  j=1,2,\cdots, s_k, \\
        H_j&=  {\dt}^{1/2} X_{j,0} + \dt^{3/2} X_{j,1}  + \cdots + \dt^{k-3/2}X_{j,k-2} , \;  j=s_k+1,\cdots, S_k.
    \end{aligned}
\end{equation}
Here, each $X_{j,q}$ is a polynomial of $H,V_j$ that satisfies $\|X_{j,q}\|=\mathcal{O}(\|\mathcal{L}\|^{q+1/2}_{\mathrm{be}})$ for $1\leq j\leq s_k$ and $\|X_{j,q}\|=\mathcal{O}(\|\mathcal{L}\|^{q+1}_{\mathrm{be}})$ otherwise.
\end{lem}

Intuitively, the unitary operator on the right-hand side of \cref{eqn:H_lemma} can be expanded and its first column can be compared to the first column of the unitary matrix in \cref{eq:lindblad_approximatios_k}. Specifically, each matrix in \cref{asymp-ham} can be obtained by matching the corresponding terms in the expansion in \eqref{asymp-kraus}.
The proof is in \cref{sec:H_existence}. According to \cref{lem:H_existence}, we obtain $\|H_j\|=\mathcal{O}(\|\mathcal{L}\|_{\mathrm{be}})$. 

Finally, to complete the proof of \cref{thm:main}, the remaining step is to demonstrate that $\widetilde{H}$ must be a Hermitian matrix. This is stated in the following lemma:
\begin{lem}\label{lem:hermitian} The dilated Hamiltonian  $\widetilde{H}$ constructed in \cref{lem:H_existence} is Hermitian.
\end{lem}
The proof of Lemma 4 is in \cref{sec:lem:hermitian}.

\section{Numerical experiments}\label{sec:nume}
In this section, we provide results from several numerical experiments to illustrate the convergence of our algorithm. We start with a time-independent transverse-field Ising model (TFIM) in \cref{sec:TFIM} and examine the convergence rate of the first-, second-, and third-order methods. The specific forms of these methods can be found in \cref{sec:all_scheme}. To extend the applications to more general cases, we also present two \emph{time-dependent} examples in \cref{sec:TFIM_dependent} and \cref{sec:spin}  to further test the performance of our proposed methods.

In all the following numerical experiments, we use the fourth-order Runge-Kutta scheme with a very small time step to generate the ``exact solution" $\rho_T$ and measure the error at time $T$ using the trace distance, which means that
\begin{equation}\label{eqn:error_def}
\mathrm{Error}=\|\rho_{N}-\rho_T\|_{1}\,,
\end{equation}
where $T$ is the stopping time, $N=T/\dt$, and $\rho_N$ is the output of our algorithm.
\subsection{A TFIM damping model}\label{sec:TFIM}
Consider the one-dimensional TFIM model defined on $m$ sites:
\begin{equation}\label{eqn:H_Ising}
H=-\left(\sum^{m-1}_{i=1} Z_{i}Z_{i+1}+Z_{L}Z_1\right) -g\sum^m_{i=1} X_i,
\end{equation}
where $g$ is the coupling coefficient that describes the transverse magnetic field strength, $Z_i$ and $X_i$ are Pauli operators for the $i$ th site and the dimension of $H$ is $2^m$. We set $m=4$ and $g=1$ and simulate the TFIM model with damping ~\cite{schlimgen2022quantum}:
\begin{equation}
\frac{d}{dt} \rho = -i[H, \rho] + \sum^J_{j=1}V_j \rho V^\dagger_j - \frac{1}{2} \left\{V^\dagger_j V_j, \rho \right\},\quad \rho(0)=\ketbra{\psi_0},
\end{equation}
where $V_j=\sqrt{\gamma}(X_j-iY_j)/2$, the damping parameter $\gamma=0.1$, and $\ket{\psi_0}$ is the ground state of $H$. In~\cite{schlimgen2022quantum}, the authors have used this model to test the accuracy of their numerical scheme and to investigate the effect of magnetic field strengths and damping parameters on the solution trajectory. For our experiment, we focus on the scaling of the error of our numerical methods with $\dt$, so we only assess its effectiveness with fixed values of $g$ and $\gamma$.

\begin{figure}
\centering
  \subfloat[Evolution of $\bra{\psi_0}\rho(t)\ket{\psi_0}$]{\includegraphics[width=7cm]{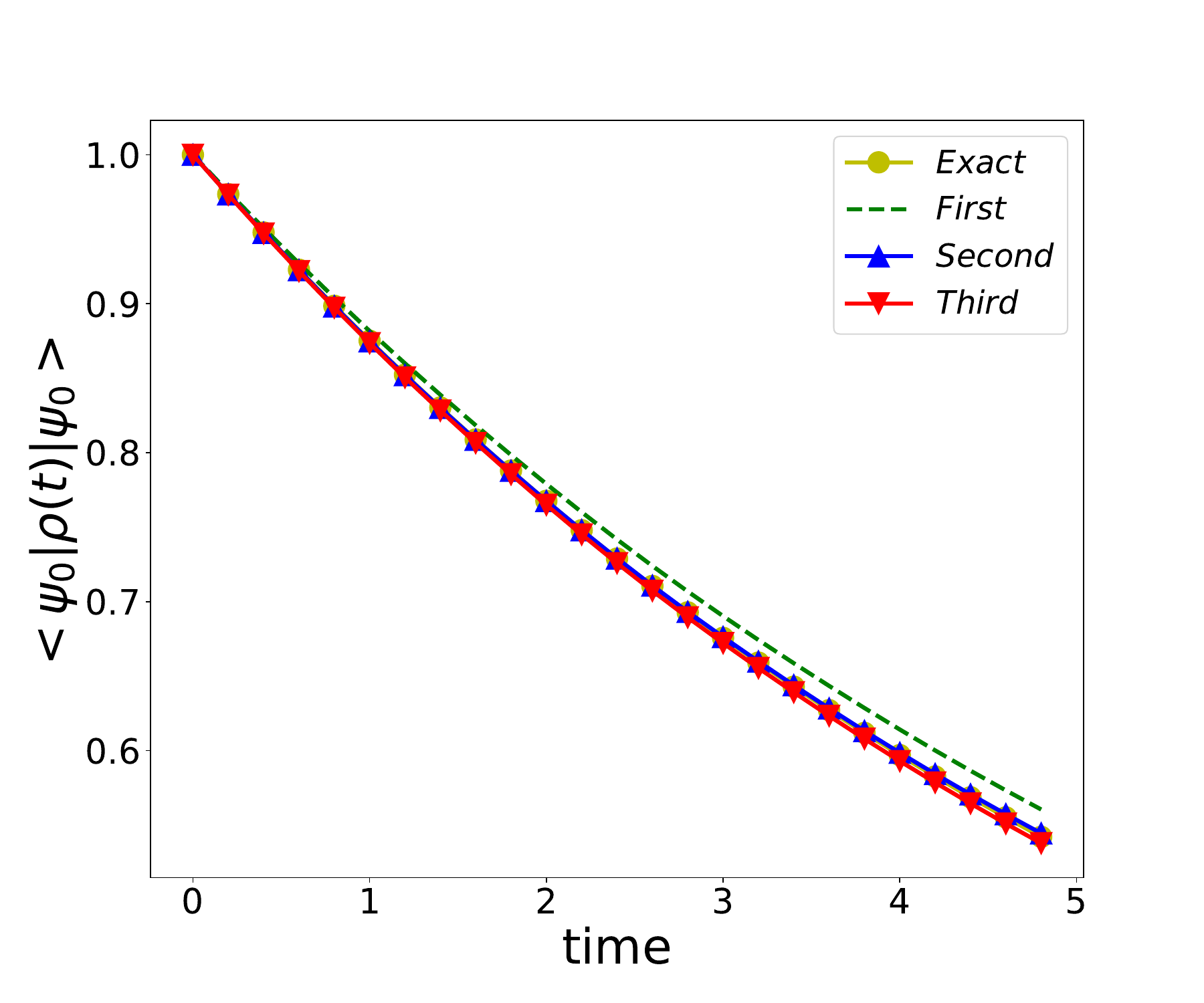}}
  \subfloat[Error at $T=1$]{\includegraphics[width=7cm]{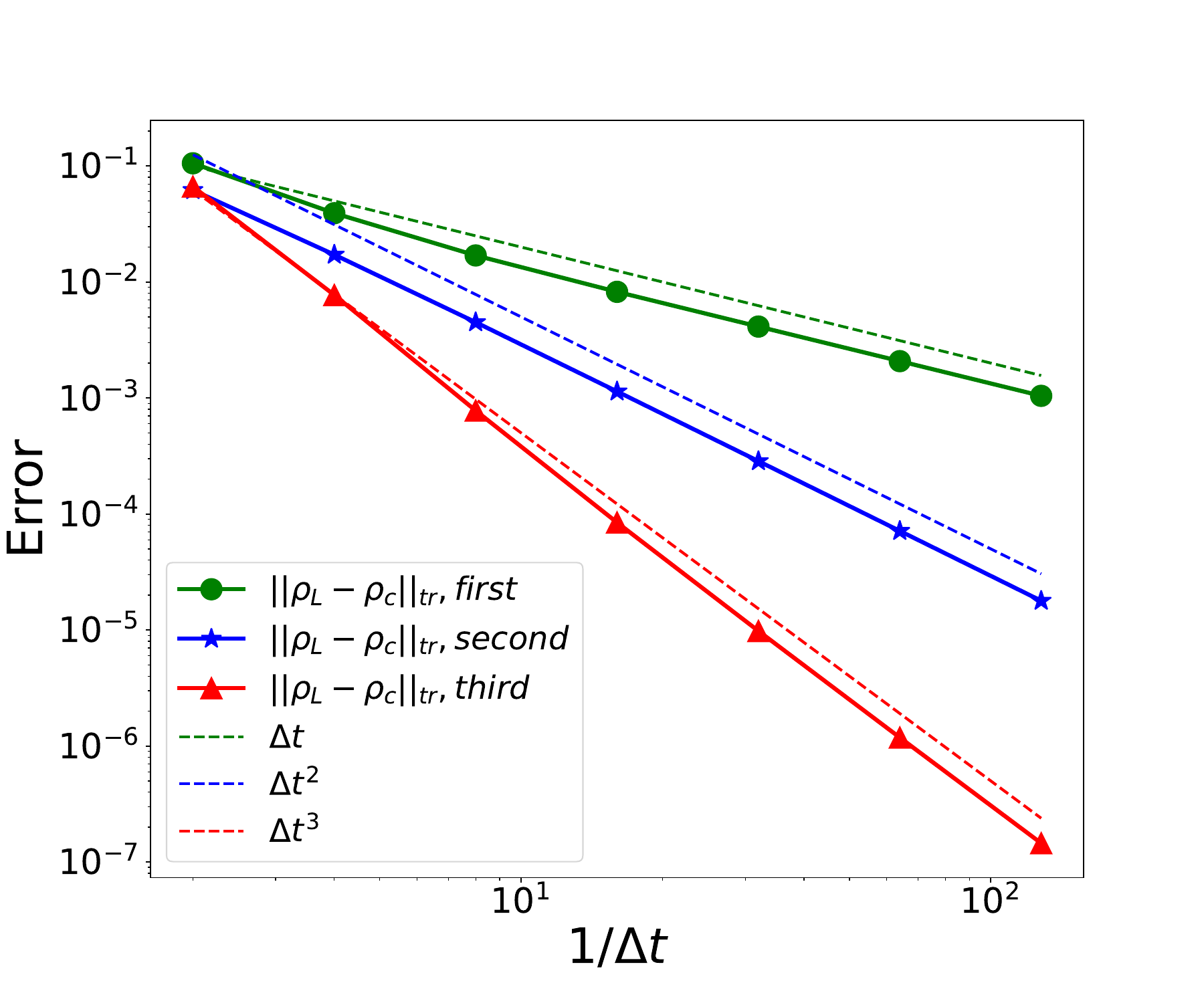}}
   \caption{Examining the accuracy of the first-, second-, and third-order methods using the TFIM damping mode~\eqref{eqn:H_Ising}. Left: The comparison of the evolution of the ground state overlap with different schemes and the same step size $\Delta t=0.1$ up to the stopping time $T=5$. Right: the comparison of the error versus $\dt$ using different schemes with stopping time $T=1$. We set the $x$ axis as $1/\Delta t$ and plot it in the log scale to illustrate the order scaling of our methods. The three dashed lines are drawn by matching the error curve from the $k$ th scheme  with $(\Delta t)^k$. The comparison of the slopes verifies that our $k$ th-order scheme indeed leads to an error that is $\Or(\Delta t)^k$.} \label{fig:1}
\end{figure}

We examine the convergence of three numerical schemes (see~\cref{sec:all_scheme}): (i) the first-order scheme in~\eqref{eqn:first_order_appendix}; (ii) the second-order scheme in~\eqref{eqn:second_order_appendix}; and (iii) the third-order scheme in~\eqref{eqn:third_order_appendix}. 
The results are shown in Figure \ref{fig:1}. The graph on the left shows the overlaps between $\rho(t)$ and the ground state when the time step $\dt=0.1$. We can see that the second- and third-order schemes match the exact solution better than the first-order scheme. In the right graph, with a stopping time of $T=1$, we have evaluated the convergence of the three methods using different $\dt$ and measured the end error using \eqref{eqn:error_def}.  One can observe that all the schemes converge in the expected order. Due to the random selection of the operators $G$ and $G_{j,2}$, as well as the initial condition, these orders of accuracy are very likely sharp.

\subsection{
A time-dependent TFIM  model with damping}\label{sec:TFIM_dependent}

In the following numerical test, we consider the time-dependent TFIM damping model, where both the Hamiltonian and jump operators are driven by a linear pulse, 
\begin{equation}\label{eqn:H_Ising_dependent}
H(t)=H+tH',\quad V_j(t)=V_{j,1}+tV_{j,2}\,.
\end{equation}
Here, $H$ is the TFIM model with $m=4,g=1$ and $H'=\frac{G+G^\dagger}{\|G+G^\dagger\|}$ with $G\sim \mathcal{N}(0,1,I_{2^m\times 2^m})$. We also choose  random damping operators 
\begin{equation}
V_{j,1}=\sqrt{\gamma}(X_j-iY_j)/2,\quad V_{j,2}=\frac{G_{j,2}}{\|G_{j,2}\|}\,,
\end{equation}
where $\gamma=0.1$ and $G_{j,2}\sim \mathcal{N}(0,1,I_{2^m\times 2^m})$. We note that this is a time-dependent Lindblad equation with two jump operators. We test the first, second, and third methods as discussed in
\cref{sec:all_scheme}. 

\begin{figure}
\centering
    \subfloat[Evolution of $\bra{\psi_0}\rho(t)\ket{\psi_0}$]{\includegraphics[width=7cm]{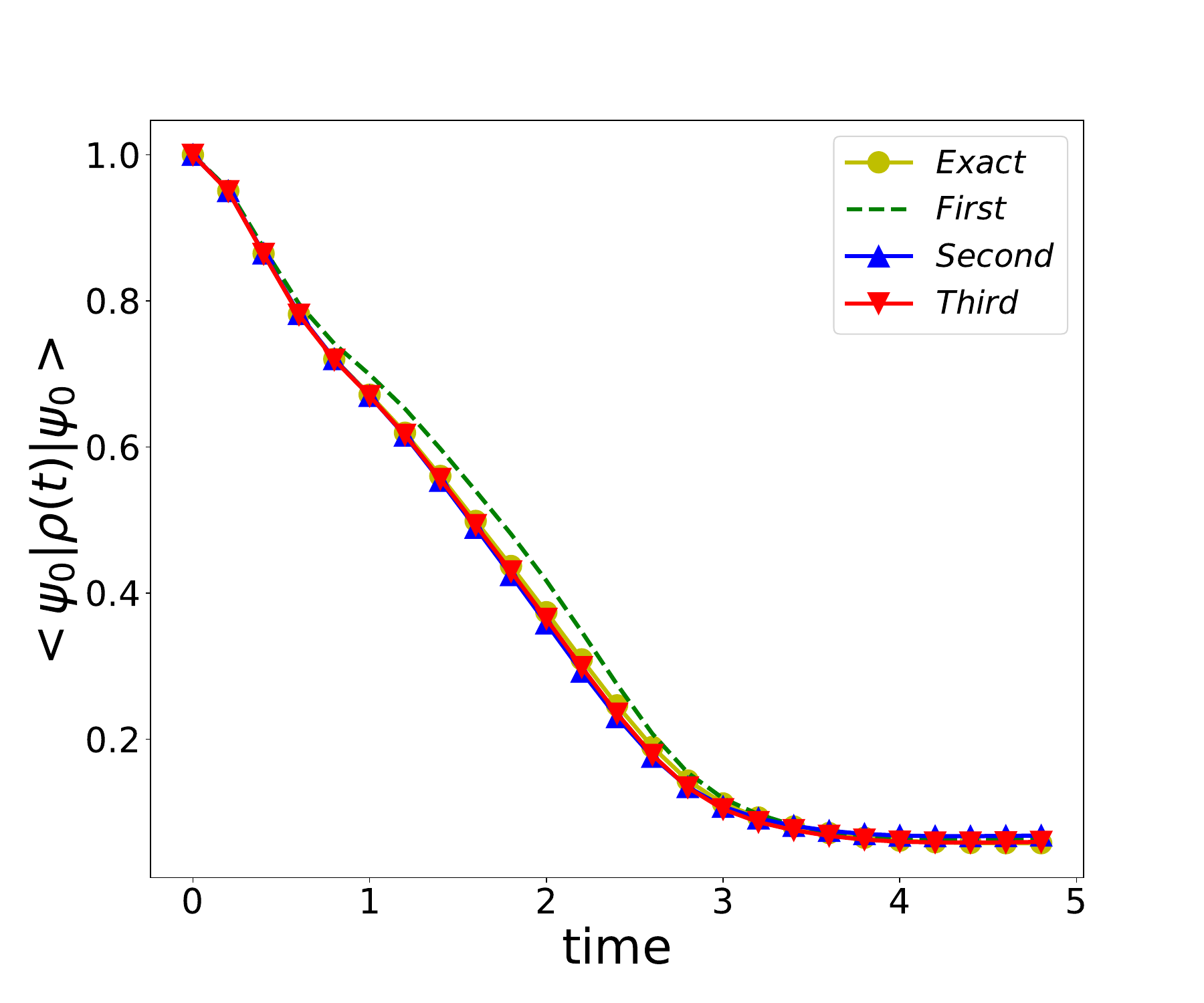}}
  \subfloat[Error at $T=1$]{\includegraphics[width=7cm]{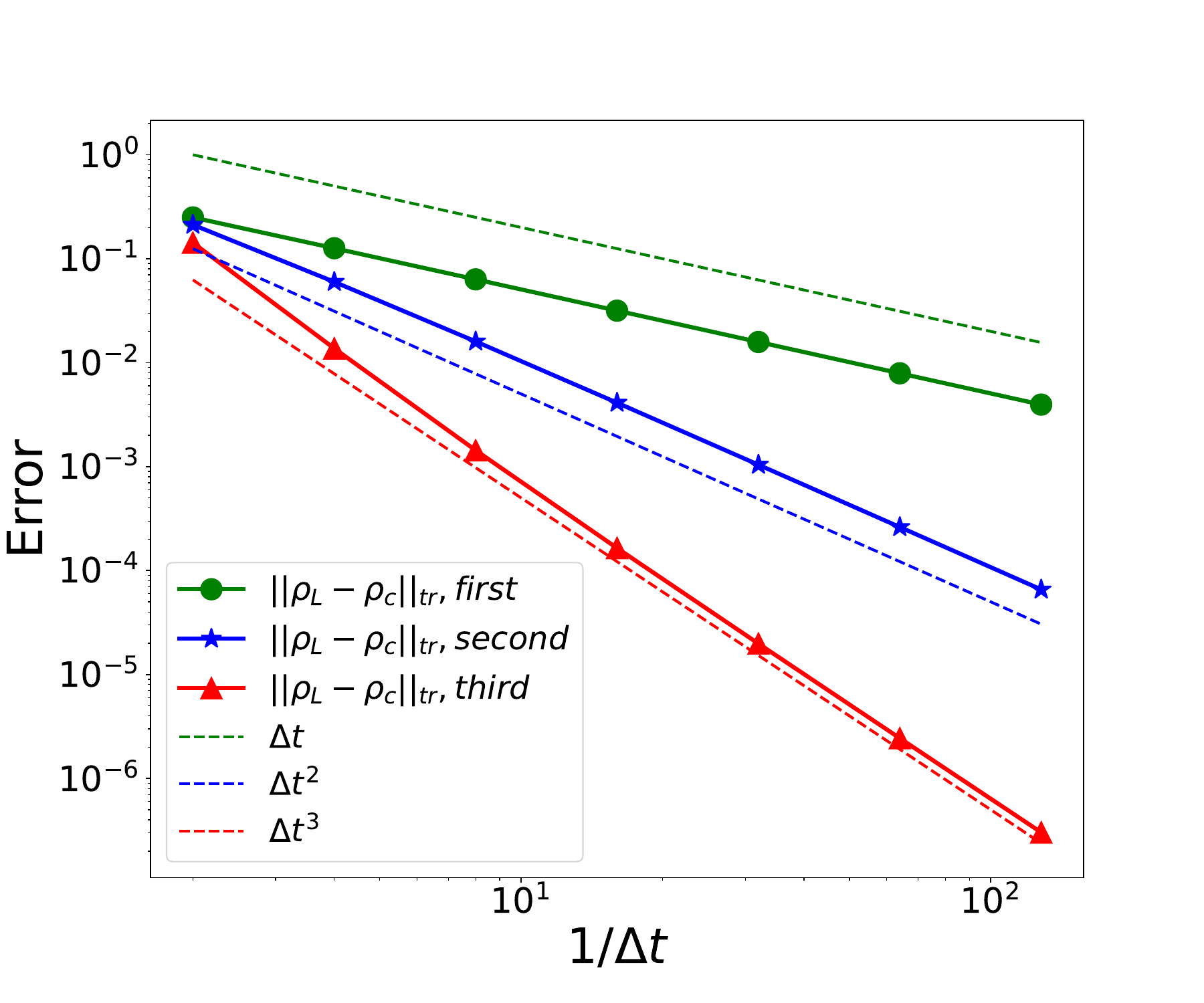}}
   \caption{Testing the accuracy of the first, second, and third order methods using the time-dependent TFIM Lindbladian~\eqref{eqn:H_Ising_dependent}. Left: The comparison of the evolution of the ground state overlap with different schemes and the same step size $\Delta t=0.1$ up to stopping time $T=5$. Right: The comparison of the error versus $\dt$ (on the logarithmic scale) using different schemes with stopping time $T=1$. We set the $x$ axis as $1/\Delta t$ and plot it in the log scale to illustrate the order scaling of our methods. The three dashed lines are drawn by matching the error curve from the $k$ th scheme with $(\Delta t)^k$. This verifies that our $k$ th-order scheme indeed leads to an error that is $\Or(\Delta t)^k$.}\label{fig:2}
\end{figure}

The result is shown in Figure \ref{fig:2}. On the left graph, we set the initial state as the ground state of $H$ and perform the simulations up to $T=5$. We compare the evolution of the overlap with the ground state for all three methods. It can be seen from the graph that the second- and third-order schemes show much better agreement with the exact solution than the first-order scheme. The results shown in the right panel are obtained with a random initial state and simulation of the dynamics up to time $T=1$. We examine the convergence of the methods by varying $\dt$ and measuring the end error as defined in \eqref{eqn:error_def}. We observe that all three schemes converge to the true solution with the expected order of accuracy.

\subsection{Periodically driven Lindbladian dynamics}\label{sec:spin}

In this section, we consider a single-qubit time-dependent system that is driven by a periodic Hamiltonian and jump operators~\cite{Scopa2019}. Specifically, we choose 
\begin{equation}
H(t)=-\frac{\sqrt{2}}{2}(1-\cos(t))\sigma_z\,.
\end{equation}
and the damping operators
\begin{equation}
V_{1}=(2+0.5\sin(t))\sigma_+,\quad V_{2}=(3-0.5\sin(t))\sigma_-\,,
\end{equation}
We then compare the performance of the first-, second-, and third-order method at the stopping time $T=10\pi$ with a random initial state. The error is measured using \eqref{eqn:error_def}. 

\begin{figure}
\centering
      \subfloat[Evolution of $\mathrm{Tr}\left(\rho(t)\sigma_z\right)$]{\includegraphics[width=7cm]{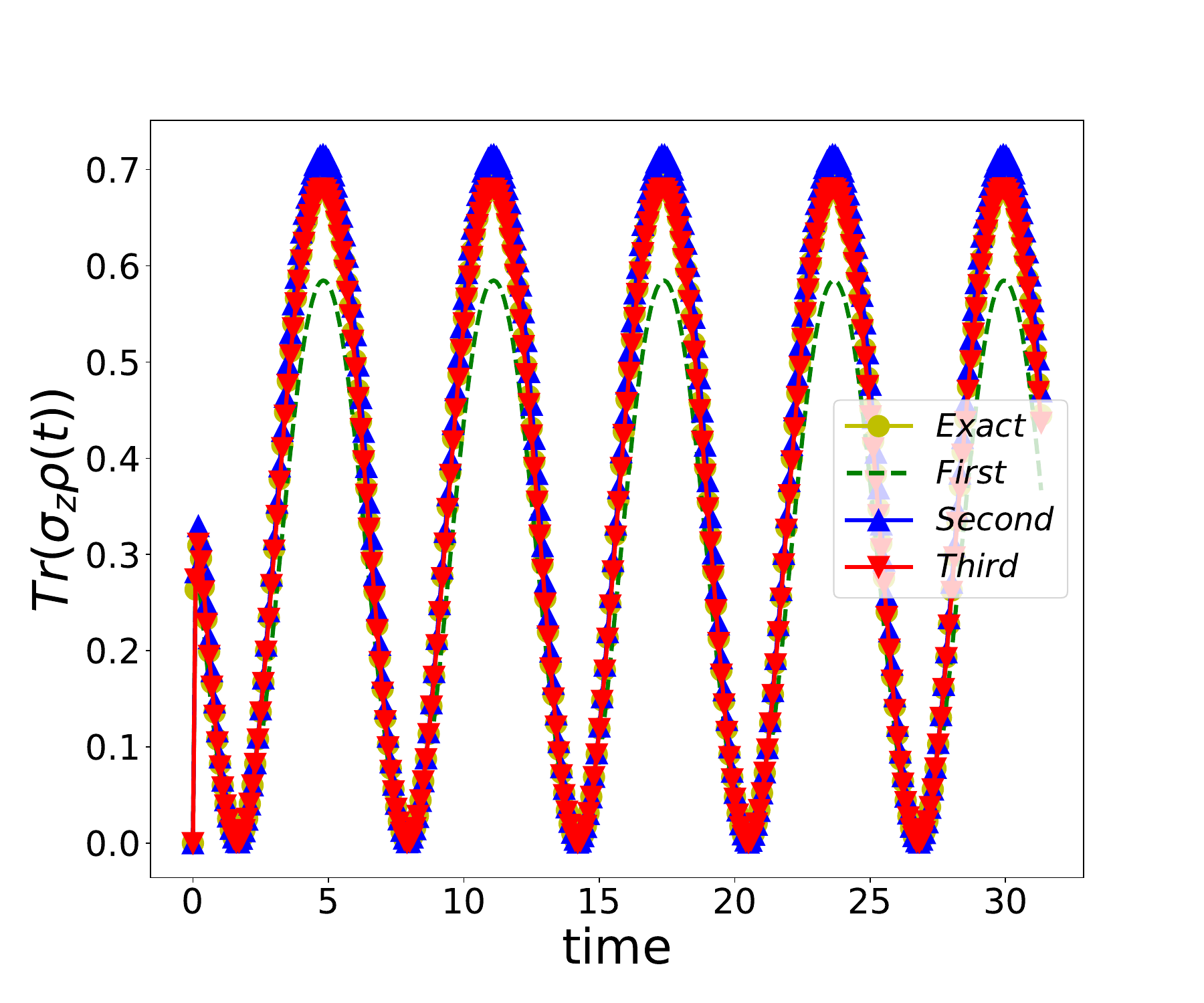}}
  \subfloat[Error at $T=10\pi$]{\includegraphics[width=7cm]{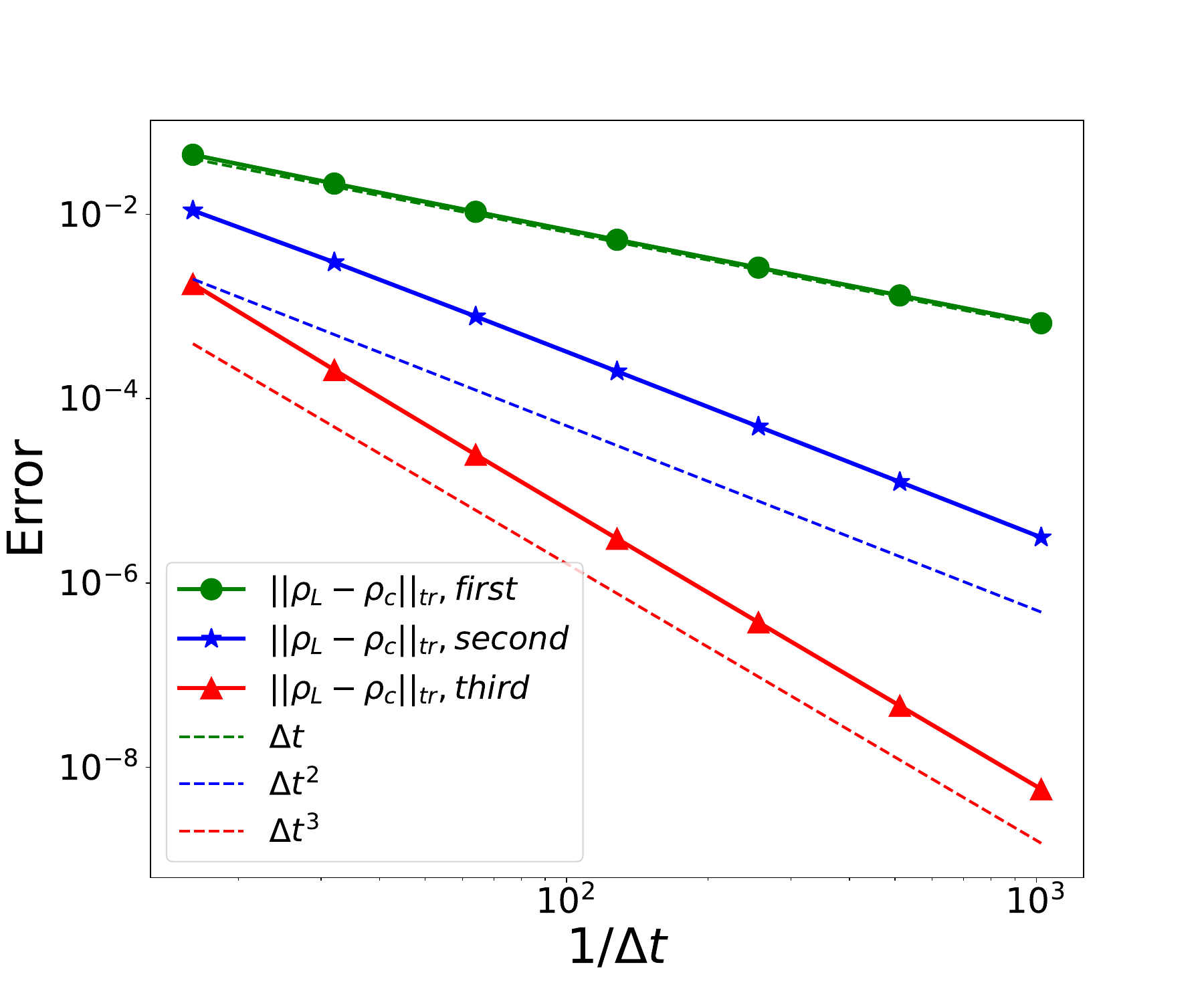}}
   \caption{Testing the accuracy of the first, second, and third order methods using the periodic driving Lindbladian~\cite{Scopa2019}. Left: we compare the evolution of $\mathrm{Tr}\left(\rho(t)\sigma_z\right)$ with different schemes and the same step size $\Delta t=0.1$ up to stopping time $T=10\pi$. Right: we compare the error vs $\dt$ using different schemes with stopping time $T=10\pi$.~\re{We set the $x$-axis as $1/\Delta t$ and plot it in the log scale to illustrate the order scaling of our methods. The three dashed lines are drawn by matching the error curve from the $k$-th scheme with $(\Delta t)^k$. The comparison of the slopes verifies that our $k$-th order scheme indeed leads to an error that is $\Or(\Delta t)^k$. }}\label{fig:3}
\end{figure}

The numerical results are summarized in Figure \ref{fig:3}. In the left graph, we choose $\dt=0.1$ and compare the evolution of $\mathrm{Tr}\left(\rho(t)\sigma_z\right)$. We observe that the second- and third-order schemes exhibit significantly better accuracy than the first-order scheme. Similar to the previous results, in the right figure, the error of all schemes behaves with the expected order of convergence.

\section{Discussion and conclusions}\label{sec:discuss}

This paper presents a new method for simulating the Lindblad dynamics using Hamiltonian simulation in an enlarged Hilbert space. Our algorithm only involves simulation of a dilated Hamiltonian and trace-out operations. The latter can be implemented simply by measuring the ancilla qubits and discarding the results. Each step of our algorithm forms a CPTP map, thereby guaranteeing a success probability of one. Contrary to previous methods~\cite{CW17,LW22}, our algorithm eliminates the need for oblivious amplitude amplification at the level of Lindbladian simulation, which may require precise adjustment of the time step $\Delta t$ with respect to the block-encoding factor.

Our methodology \re{bridges the gap between Lindblad simulation and Hamiltonian simulation.} This approach  also introduces a new class of Hamiltonian simulation problems, where the Hamiltonian $\widetilde{H}$ consists of commutators among the jump operators (including the system Hamiltonian $H$) and the various components of the dilated Hamiltonian scale differently with respect to the time step $\Delta t$. Identifying suitable Hamiltonian simulation techniques for this specific context poses an interesting question for future investigations.  For example, suppose that both $H$ and $V_j$ can be expressed as sums of Pauli operators. In that case, we can decompose $H_0, H_1, \ldots, H_{S_k}$ into sums of Pauli operators and further refine the simulation using a high-order Trotterization method.

In contrast to Hamiltonian simulations, where a diverse range of methods are available and practicality resource estimates have been conducted (see e.g., \cite{childs2018toward}), quantum algorithms for Lindblad simulations remain in their nascent stages. This study introduces a framework that differs from those in the existing literature.  Low-order methods, such as second and third order, may be more practical than higher-order versions in terms of practical implementation. We hope that this work can facilitate future resource estimates for identifying the most practical methods for simulating Lindblad dynamics.

\section*{Acknowledgements}   This material is based upon work supported by the U.S. Department of Energy, Office of Science,
Q11 National Quantum Information Science Research Centers, Quantum Systems Accelerator (ZD). Additional funding is provided by the Challenge Institute for Quantum Computation (CIQC) funded by the National Science Foundation (NSF) through Grant No. OMA-2016245 and a Google Quantum Research Award (LL). L.L. is a Simons investigator in Mathematics. X.L.'s research is supported by the NSF Grants No. DMS-2111221 and No. CCF-2312456. Z.D. and L.L. thank the Institute for Pure and Applied Mathematics
(IPAM) for its hospitality in hosting them as long-term visitors during the semester-long program   ``Mathematical and Computational Challenges in Quantum Computing'' in Fall 2023.




\bibliography{ref}

\appendix 

\section{Derivation of the time-independent second-order scheme}\label{sec:second_order_scheme}
As a concrete example, in this appendix, we derive a second-order scheme to simulate time-independent Lindbladian dynamics. 

\textbf{Step 1: Formulate the weak scheme of order two for the SDE \eqref{eqn:SDE_for_Lindblad}.} 

 According to the first step of \cref{alg:c_H}, we can write down the weak order-2.0 scheme according to~\cite[(14.2.6)]{kloeden1992numerical}:
\[
\begin{aligned}
    L_{2,\dt}[\ket{\psi}]=&\left(\ket{\psi_n} + V_0 \ket{\psi_n} \dt + \frac{1}{2} V^2_0\ket{\psi_n}\dt^2\right)\\
+&\sum^J_{j=1}\left( V_j\int^{\Delta t}_0 \mathrm{d}W^j_{s_1}
+V_jV_0\int^{\Delta t}_{0}\int^{s_2}_0 \mathrm{d}s_1 \mathrm{d}W^j_{s_2}+V_0V_j\int^{\Delta t}_0\int^{s_2}_0 \mathrm{d}W^j_{s_1} \mathrm{d}s_2\right)\\
+&\sum^J_{j,k=1}\int^{\Delta t}_0\int^{s_2}_0 \mathrm{d}W^k_{s_1} \mathrm{d}W^k_{s_2}
\end{aligned}
\]
Here we have defined,
\begin{equation}\label{op-V0}
    V_0=-iH-\frac12 \sum^J_{j=1}V^\dag _jV_j.
\end{equation}

\textbf{Step 2: Formulate the second-order Kraus form.}

In the second step, we construct the Kraus form according to the scheme described above.  Generally, we must convert the It\^o integrals to random variables and arrange them to ensure that they are not correlated (see, e.g., Lemma \ref{lem:R_alpha}). In this case, we simply take the formula from~\cite[(14.2.7)]{kloeden1992numerical} and reformulate the above second-order scheme as follows:
\begin{equation}\label{sde-2ndorder}
    \begin{aligned}
    \ket{\psi_{n+1}} = &\left(\ket{\psi_n} + V_0 \ket{\psi_n} \dt + \frac{1}{2} V^2_0\ket{\psi_n}\dt^2\right) \\
    &+\sum^J_{j=1}\left(V_j + \frac{\dt}{2} (V_j V_0 + V_0V_j) \right) \ket{\psi_n} \dw_j\\
    & + \frac12 \sum^J_{j=1}V_j^2  \ket{\psi_n} ( \dw_j^2 - \dt) \\
 & + \frac12 \sum^J_{j_1\neq j_2}V_{j_2}V_{j_1}  \ket{\psi_n} (\dw_{j_1}\dw_{j_2} - \dz_{j_1,j_2}).
\end{aligned}
\end{equation}
Here, $\{\dw_j\}^J_{j=1}$ are independent Gaussian random variables
with mean zero and variance $\dt$, and $\{\dz_{j_1,j_2}\}$ are independent two-point random variables such that 

\[
\mathbb{E}(\dz_{j_1,j_2})=0,\quad \mathbb{E}(|\dz_{j_1,j_2}|^2)=\dt^2\,,
\]
for $j_2=1,2,\cdots,j_1-1$ and $\dz_{j_1,j_2}=-\dz_{j_2,j_1}$.

Given that the random noises in distinct terms are uncorrelated, and taking the expectation on both sides, we arrive at the following relation for the expected state at time $n+1$
\[
\mathbb{E}(\ketbra{\psi_{n+1}})=F_0 \mathbb{E}(\ketbra{\psi_{n}}) F_0^\dag + \sum^J_{j=1}F_{1,j}  \mathbb{E}(\ketbra{\psi_{n}}) F_{1,j}^\dag + \sum^J_{j,k}F_{2,j,k}\mathbb{E}(\ketbra{\psi_{n}}) F_{2,j,k}^\dag\,,
\]
where
\begin{align*}
    F_0=& I + V_0 \dt + \frac12 V_0^2 \dt^2, \\
    F_{1,j}=&-i \sqrt{\dt} \left( V_j + \frac{\dt}{2} (V_j V_0 + V_0V_j)\right),\quad\forall 1\leq j\leq J\,,\\
    F_{2,j,k} = &  -i\frac{\sqrt{2} \dt}{2} V_jV_k,\quad \forall 1\leq j,k\leq J. 
\end{align*}
Here, we have combined the third and fourth lines of \eqref{sde-2ndorder} in $F_{2,j,k}$ using $\mathbb{E}(( \dw_j^2 - \dt)^2)=2\dt^2$ and $\mathbb{E}((\dw_{j_1}\dw_{j_2} - \dz_{j_1,j_2})^2)=2\dt^2$. This leads us to define the Kraus form
\[
\mathcal{K}[\rho]=F_0 \rho F_0^\dag + \sum^J_{j=1}F_{1,j}\rho F_{1,j}^\dag + \sum^J_{j,k}F_{2,j,k}\rho F_{2,j,k}^\dag,
\]
and to define the iteration scheme as
\[
\rho_{n+1}=\mathcal{K}[\rho_n]\,.
\]

\textbf{Step 3: Construct the dilated Hamiltonian $\widetilde{H}$.}

The goal of the last step is to construct the Hamiltonian $\widetilde{H}$ such that
\begin{equation}\label{dilate-2ndorder}
\mathcal{K}[\rho]=\mathrm{Tr}_A\left(\exp(-i\sqrt{\dt}\widetilde{H})\left(\ketbra{0}\otimes\rho_n\right)\exp(i\sqrt{\dt}\widetilde{H})\right)+\mathcal{O}(\dt^3)
\end{equation}
Since there are $J^2+J+1$ Kraus operators, we seek a Hamiltonian with the following block structure,
\[
\widetilde{H}= \begin{bmatrix}
    H_0 & \cdots&H_{1,j}^\dag & \cdots&H_{2,j,k}^\dag \\
    \cdots& 0&0&0&0\\
    H_{1,j} & 0&0&0&0\\
    \cdots& 0&0&0&0\\
    H_{2,j,k} & 0&0&0&0
\end{bmatrix}\,,
\]
where we require $H_0$ to be a Hermitian matrix.

We begin by noting that 
\[
\begin{aligned}
&\mathrm{Tr}_A\left(\exp(-i\sqrt{\dt}\widetilde{H})\ketbra{0}\otimes\rho\exp(i\sqrt{\dt}\widetilde{H})\right)\\
=&\mathrm{Tr}_A\left(\left(\exp(-i\sqrt{\dt}\widetilde{H})\ket{0}\otimes I_n\right) I_A\otimes \rho \left(\bra{0}\otimes I_n\exp(i\sqrt{\dt}\widetilde{H})\right)\right)\\
=&\sum^{J^2+J+1}_{j=1}\left(\bra{j}\otimes I_n\exp(-i\sqrt{\dt}\widetilde{H})\ket{0}\otimes I_n\right) I_A\otimes \rho \left(\bra{0}\otimes I_n\exp(i\sqrt{\dt}\widetilde{H})\ket{j}\otimes I_n\right)\,.
\end{aligned}
\]
This will be compared to the Stinespring form,
\[
\begin{aligned}
    \mathcal{K}[\rho]=&\mathrm{Tr}_A\left(\begin{bmatrix}
        F_0 & \vdot & \cdots & \vdot\\
         \vdot & \vdot & \cdots & \vdot\\
        F_{1,j} & \vdot & \cdots & \vdot\\
         \vdot & \vdot & \cdots & \vdot\\
        F_{2,j,k} & \vdot & \cdots & \vdot\;
    \end{bmatrix}\right.\ketbra{0}\otimes \rho \left.\begin{bmatrix}
        F_0 & \vdot & \cdots & \vdot\\
         \vdot & \vdot & \cdots & \vdot\\
        F_{1,j} & \vdot & \cdots & \vdot\\
         \vdot & \vdot & \cdots & \vdot\\
        F_{2,j,k} & \vdot & \cdots & \vdot\;
    \end{bmatrix}^\dagger\right)\\
    =&\sum^{J^2+J+1}_{j=1}\left(\bra{j}\otimes I_n \left(\ketbra{j}{0}\otimes F_j\right)\ket{0}\otimes I_n\right) I_A\otimes \rho \left(\bra{0}\otimes I_n \left(\ketbra{0}{j}\otimes F^\dagger_j\right)\ket{j}\otimes I_n\right)\,.
\end{aligned}
\]
By matching the above two equations, we see that, to arrive at
\eqref{dilate-2ndorder}, we need to find $H_0,H_{1,j}$, and $H_{2,j,k}$ so that
\begin{equation}\label{eqn:matching_second_order}
    \bra{[\cdot]}\exp(-i\sqrt{\dt}\widetilde{H}) \ket{0}=F_{[\cdot]}+\mathcal{O}((\dt)^{3})\,.
\end{equation}
where $[\cdot]$ is $0$, $(1,j)$, or $(2,j,k)$.

Next, we note that the matrix exponential can be expanded in the following Taylor expansion
\[
  e^{-i \sqrt{\dt} \widetilde{H}} =  I - i\dt^{1/2}\widetilde{H} - \frac{\dt}{2} \widetilde{H}^2 + \frac{i\dt^{3/2}}{6}\widetilde{H}^3 + \frac{\dt^2}{24} \widetilde{H}^4+ \cdots
\]
Plugging this formula into the left-hand side of \eqref{eqn:matching_second_order}, we match terms in the blocks of the first column and find that 
\begin{equation}\label{eqn:mathcing_second_order}
\begin{aligned}
    F_0 = & I-i\sqrt{\dt}H_0-\frac{\dt}{2}(H^2_0+Q)+i\frac{\dt^{3/2}}{6}\left(H^3_0+H_0Q+QH_0\right)\\
    &+ \frac{\dt^2}{24}\left(H^4_0+Q^2+H^2_0Q+QH^2_0+H_0QH_0\right)+\co(\dt^{5/2}) \\
    F_{1,j} =&  -i\sqrt{\dt} H_{1,j}-\frac{1}{2}\dt H_{1,j}H_0 +i \frac{\sqrt{\dt}\dt }{6} H_{1,j} (Q +H^2_0) + \co(\dt^{5/2}),\quad \forall 1\leq j\leq J\,,\\ 
     F_{2,j,k} =&  -i\sqrt{\dt} H_{2,j,k}-\frac{1}{2}\dt H_{2,j,k}H_0 +i\frac{\sqrt{\dt}\dt }{6} H_{2,j,k} (Q +H^2_0) + \co(\dt^{5/2}),\quad \forall 1\leq j,k\leq J\,,\\ 
\end{aligned}
\end{equation}
where 
\[
Q=\sum^J_{j=1}H^\dagger_{1,j}H_{1,j}+\sum^J_{j,k=1}H^\dagger_{2,j,k}H^\dagger_{2,j,k}\,.
\]
We first match the first-order terms in the last two equations by taking the leading terms to obtain 
\[
\begin{aligned}
    H_{1,j} = V_{j} + \co(\dt)=:X_{1,j,0}+\co(\dt),\quad H_{2,j,k} =\frac{\dt^{1/2}}{\sqrt{2}}V_{j}V_{k} + \co(\dt^{3/2})=:\dt^{1/2}X_{2,j,k,0}+\co(\dt^{3/2}). 
\end{aligned}
\]
where we use $X_{[\cdot]}$ to represent the coefficient of the order terms $\dt^{p}$.

We then substitute them into $Q$ to obtain
\[
Q=\sum^J_{j=1}V^\dagger_jV_j+\mathcal{O}(\Delta t)=:Q_{0}+\mathcal{O}(\Delta t)\,.
\]
Plugging this into the first equation of \eqref{eqn:mathcing_second_order} and matching the first term, we obtain
\[
H_0=i\dt^{1/2}\left(V_0+\frac{1}{2}Q_0\right)=\dt^{1/2}H + \co(\dt^{3/2})\,.
\]
Next, we include the next-order terms in $F_{1,j}$ and $ F_{2,j,k}$. Again, matching both sides of the last two equations, we obtain the asymptotic expansion,
\[
\begin{aligned}
    H_{1,j} = &X_{1,j,0}+\dt X_{1,j,1}=: V_j + \dt \left(\frac12 V_0V_j + \frac12V_jV_0+\frac16V_j\sum^J_{j'=1}V^\dag_{j'}V_{j'}+\frac{i}{2}V_jH\right),\\
    H_{2,j,k} =& \dt^{1/2}X_{2,j,k,0}+\dt^{3/2} X_{2,j,k,1}=:\frac{\dt^{1/2}}{\sqrt{2}}  V_{j}V_k+\dt^{3/2}\left(\frac{1}{6\sqrt{2}}V_{j}V_k\sum^J_{j'=1}V^\dag_{j'}V_{j'}+\frac{i}{2\sqrt{2}}V_{j}V_kH\right). 
\end{aligned}
\]
We then substitute them into $Q$ and obtain
\[
\begin{aligned}
    Q=&\sum^J_{j=1}V^\dagger_jV_j+\dt\sum^J_{j=1} \left(V^\dagger_j V_0V_j+\frac{1}{2}V^\dagger_jV_jV_0+\frac{1}{2}V_0V^\dagger_jV_j+\frac{1}{3}\left(\sum^J_{j'=1}V^\dag_{j'}V_{j'}\right)^2+\frac{i}{2}\left(V^\dagger_jV_jH-HV^\dagger_jV_j\right)\right)\\
    &+\mathcal{O}(\dt^2)\\
    =:&Q_{0}+\Delta t Q_1+\mathcal{O}(\dt^2)
\end{aligned}
\]
Plugging this into the first equation of \eqref{eqn:mathcing_second_order} and matching the second term, we find that

\[
H_0=\dt^{1/2} X_{0,1}+\dt^{3/2}X_{0,2}\,,
\]
where 
\[
\begin{aligned}
    X_{0,2}=&\frac{i}{2}\left(V^2_0+X^2_{0,1}+Q_1\right)+\frac{1}{6}\left\{X_{0,1},Q_0\right\}-\frac{i}{24}Q^2_0\\
    =&-\frac{1}{12}\left\{H,\sum^J_{j=1}V^\dag_jV_j\right\}.
\end{aligned}
\]
We have left out higher-order terms, since they only contribute at most $\co(\dt^{3})$ terms, which is comparable to the leading error term in \cref{dilate-2ndorder}.
This completes the construction of the Hamiltonian $\widetilde{H}$. 

\section{A summary of first-, second-, and third-order schemes for simulating time-dependent Lindblad equations}\label{sec:all_scheme}
In this appendix, we extend the construction in the previous appendix and derive the numerical schemes to the \emph{time-dependent} Lindblad equation, which takes the form:
\begin{equation}\label{eq:lindblad_time_dependent}
\frac{\rd\rho}{\rd t}=\mathcal{L}_t(\rho)=:-i[H(t),\rho]+\sum^J_{j=1}V_j(t)\rho V^\dagger_j(t)-\frac{1}{2}\left\{V^\dagger_j(t)V_j(t),\rho\right\}\,.
\end{equation} 
In our derivation, we assume that $H(t),V_j(t)\in\mathbb{C}^2[0,\infty)$. We note that when $H(t),V_j(t)$ are smooth enough, we can directly implement the strategy (\cref{alg:c_H}) in this paper to develop a high-order scheme. In this appendix, we summarize the first-, second-, and third-order schemes for solving the time-dependent Lindblad equation in \cref{eq:lindblad_time_dependent}.
The scheme for solving time-independent Lindblad equations can readily be obtained by removing the terms involving the time derivatives of $H$ and $V$.

We define $V_0(t)=-iH(t)-\frac12\sum^J_{j=1}V^\dag_j(t)V(t)$. For simplicity, we omit the first step and start by expressing the Kraus operators in an asymptotic form (we omit $-i$ in front of $F$ for simplicity since it does not affect the Kraus representation), 
\begin{equation}
\begin{aligned}
F_0=  &I + V_0 \dt + \frac12 (V_0^2+V_0') \dt^2 + \frac16 (V_0^3+(V_0^2)'+V_0'V_0+V_0'') \dt^3, \\
=:&I+Y_{0,0}\dt + Y_{0,1}\dt^2 + Y_{0,2}\dt^3\\
F_{1,j}=  &\dt^{1/2} V_j + \frac{\dt^{3/2}}{2} (V'_j+V_jV_0+V_0V_j) \\
&+ \frac{\dt^{5/2}}{6} \left(V_0^2V_j+V_0'V_j+V_0V_jV_0+(V_0V_j)'+V_jV_0^2+V'_jV_0+(V_jV_0)'+V''_j\right) ,\quad \forall 1\leq j\leq J\,,\\
=:&Y_{1,j,0}\dt^{1/2} + Y_{1,j,1}\dt^{3/2} + Y_{1,j,2}\dt^{5/2}\\
F_{2,j}=  &\frac{\sqrt{\dt}\dt}{\sqrt{12}}\left(V_0V_j-V_jV_0-V'_j\right)=:Y_{2,j,1}\dt^{3/2},\quad \forall 1\leq j\leq J\,, \\
F_{3,j,k,l}=&\frac{\sqrt{\dt}\dt}{\sqrt{6}}V_jV_kV_l=: Y_{2,j,k,l,1}\dt^{3/2},\quad \forall 1\leq j,k,l\leq J\,,\\
F_{4,j,k}=&\sqrt{2} \dt \left( \frac12 V_jV_k+\frac{\dt}{6} \left(V_0V_jV_k+V_jV_0V_k + V'_jV_k+V_jV_kV_0+(V_jV_k)'\right) \right),\quad \forall 1\leq j,k\leq J\,.\\
=:&Y_{4,j,k,0}\dt + Y_{4,j,k,1}\dt^{2}
\end{aligned}
\end{equation}
Here, $Y_{[\cdot]}$ contains the coefficient of the order term $\dt^p$ in each expansion.

We note that, in the time-independent case, all derivative terms with $'$ and $''$ are equal to zero. After obtaining the above formula, we can use our general strategy in \cref{sec:pf_thm} to derive $\widetilde{H}$. For simplicity, we omit the derivation process and directly give the formulas of different order schemes:~
\begin{itemize}
\item \textbf{The first order scheme:} $\widetilde{H}=\ketbra{0}\otimes H_0+\sum^{J}_{j=1}\left(\ketbra{j}{0}\otimes H_{1,j}+\ketbra{0}{j}\otimes H^\dagger_{1,j}\right)$,  where
    \begin{equation}\label{eqn:first_order_appendix}
    H_0=\dt^{1/2}\left(iY_{0,0}+i\frac{Q_0}{2}\right)=:\dt^{1/2}X_{0,0},\quad H_{j,0}=Y_{1,j,0}=:X_{1,j,0}
    \end{equation}
    with $Q_0=\sum_{j=1}^{J} Y_{1,j,0}^\dag Y_{1,j,0}$ for all $1\leq j\leq J$. 
    
    Direct calculations yield, 
    \begin{equation}\label{formulas-X00-X1j0-Q0}
        \begin{aligned}
            X_{0,0} =& H, \\
            X_{1,j,0} = & V_j \\
            Q_0= & \sum_{j=1}^J V_j^\dag V_j.
        \end{aligned}
    \end{equation}
    Altogether,
    the dilated Hamiltonian is given by,
    \begin{align}
        \widetilde{H}=
        \begin{bmatrix}
             \sqrt{\dt} H & V_1^\dag &  V_2^\dag & \cdots & V_J^\dag \\
             V_1 & 0 & 0 & \cdots & 0 \\
             V_2 & 0 & 0 & \cdots & 0 \\
             \vdots & \vdots & \vdots & \ddots & \vdots \\
             V_J & 0 & 0 & \cdots & 0 \\
        \end{bmatrix},
    \end{align}
which is a direct generalization of \eqref{eqn:update_first_order}.
    
\item \textbf{The second-order scheme: }
\begin{equation}\label{eqn:second_order_appendix}
\begin{aligned}
    \widetilde{H}=&\ketbra{0}\otimes H_0+\sum^{J}_{j=1}\left(\left(\ketbra{j}{0}\otimes  H_{1,j}+\ketbra{0}{j}\otimes H^\dagger_{1,j}\right)+\left(\ketbra{j+J}{0}\otimes H_{2,j}+\ketbra{0}{j+J}\otimes H^\dagger_{2,j}\right)\right)\\
    &+\sum^{J}_{j,k,l=1}\ketbra{j+kJ+lJ^2-J^2+J}{0}\otimes H_{3,j,k,l}+\ketbra{0}{j+kJ+lJ^2-J^2+J}\otimes H^\dagger_{3,j,k,l}\\
    &+\sum^{J}_{j,k=1}\ketbra{j+kJ+J^3+J}{0}\otimes H_{4,j,k}+\ketbra{0}{j+kJ+J^3+J}\otimes H^\dagger_{4,j,k}\\
\end{aligned}.
\end{equation}

 Using $X_{0,0},X_{1,j,0},Q_0$ in \cref{formulas-X00-X1j0-Q0} from the first order scheme, we have the expressions for the entries of $\widetilde{H}$, $j,k,l \in [J],$
\begin{equation}
    \left\{
    \begin{aligned}
        H_{1,j}&=X_{1,j,0}+\dt\left(Y_{1,j,1} - X_{1,j,0} Z_1\right)=:X_{1,j,0}+\dt X_{1,j,1}\,, \\
        H_{2,j}&=\dt Y_{2,j,1}=:\dt X_{2,j,1}\,, \\
        H_{3,j,k,l}&=\dt Y_{3,j,k,l,1}=:\dt X_{3,j,k,l,1}\,, \\
H_{4,j,k}&=\dt^{1/2}Y_{4,j,k,0}=:\dt^{1/2}X_{4,j,k,0}\,,\\
    \end{aligned}
    \right.
\end{equation}
where 
\begin{equation}
Z_1=-\frac{i}{2} X_{0,0} - \frac16 Q_0 \,.
\end{equation}
In addition, the first diagonal block is given by,
\[
H_0=\dt^{1/2}\left(iY_{0,0}+i\frac{Q_0}{2}\right)+\dt^{3/2}\left(iY_{0,1}+\frac{i}{2}(Q_1+X^2_{0,0})-\frac{i}{24}Q^2_0+\frac{1}{6}\{Q_0,X_{0,0}\}\right)\,,
\]
where 
\begin{equation}
Q_1=\sum_{j=1}^{J} \big( X_{1,j,0}^\dag X_{1,j,1} + X_{1,j,1}^\dag X_{1,j,0} \big) + \sum_{j,k} X_{4,j,k,0}^\dag X_{4,j,k,0}\,.
\end{equation}

We find the explicit form of $\widetilde{H}$~
\begin{equation}\label{eqn:second_order_appendix'}
\begin{aligned}
    \widetilde{H}=&\ketbra{0}\otimes \left(\sqrt{\dt}H+\dt^{3/2}\left(\frac{1}{2}H'-\frac{1}{12}\left\{H,\sum V^\dagger_jV_j\right\}\right)\right)\\
    &+\sum^{J}_{j=1} \left(\ketbra{j}{0}\otimes \left(V_j + \frac{\dt}{2}\left( \{V_j,V_0\}+V_j'+\frac{1}{6}V_j\left(\sum V_j^\dagger V_j\right)+\frac{i}{2}V_jH\right)\right)\right.\\
    &\left.+\ketbra{0}{j}\otimes \left(V_j + \frac{\dt}{2} \left(\{V_j,V_0\}+V_j'+\frac{1}{6}V_j\left(\sum V_j^\dagger V_j\right)+\frac{i}{2}V_jH\right)\right)^\dagger\right) \\ 
    &+ \frac{\dt}{\sqrt{12}} \sum^{J}_{j=1} \left(\ketbra{j+J}{0} \otimes  \big([V_0, V_j] -V_j'\big)+\ketbra{0}{j+J} \otimes \big([V_0, V_j] -V_j'\big)^\dagger\right)\\
    &+\frac{\dt}{\sqrt{6}}\sum^{J}_{j,k,l=1}\ketbra{j+kJ+lJ^2-J^2+J}{0} \otimes  V_j V_k V_l +\ketbra{0}{j+kJ+lJ^2-J^2+J}\otimes  (V_j V_k V_l)^\dagger\\
    &+\sqrt{\frac{\dt}{2}}\sum^{J}_{j,k=1}\ketbra{j+kJ+J^3+J}{0} \otimes  V_j V_k+\ketbra{0}{j+kJ+J^3+J} \otimes  (V_j V_k)^\dag.\\
\end{aligned}
\end{equation}

\item \textbf{The third-order scheme:}
\begin{equation}\label{eqn:third_order_appendix}
\begin{aligned}
    \widetilde{H}=&\ketbra{0}\otimes H_0+\sum^{J}_{j=1}\left(\left(\ketbra{j}{0} \otimes  H_{1,j}+\ketbra{0}{j} \otimes H^\dagger_{1,j}\right)+\left(\ketbra{j+J}{0} \otimes  H_{2,j}+\ketbra{0}{j+J} \otimes  H^\dagger_{2,j}\right)\right)\\
    &+\sum^{J}_{j,k,l=1}\ketbra{j+kJ+lJ^2-J^2+J}{0} \otimes H_{3,j,k,l}+\ketbra{0}{j+kJ+lJ^2-J^2+J} \otimes  H^\dagger_{3,j,k,l}\\
    &+\sum^{J}_{j,k=1}\ketbra{j+kJ+J^3+J}{0} \otimes  H_{4,j,k}+\ketbra{0}{j+kJ+J^3+J} \otimes  H^\dagger_{4,j,k}\\
\end{aligned}\,,
\end{equation}
Define $X_{[\cdots],0,1},Q_0,Q_1,Z_1$ as in the first- and second-order schemes, we have
\[
H_{1,j}=Y_{1,j,0}+\dt\left(Y_{1,j,1} - X_{1,j,0} Z_1\right)+\dt^2\left(Y_{1,j,2} - X_{1,j,1} Z_1 - X_{1,j,0} Z_2\right)
=:\cdots+\dt^2 X_{1,j,2}\,,
\]
\[
H_{2,j}=\dt Y_{2,j,1}+\dt^2\left(Y_{2,j,2} - X_{2,j,1} Z_1\right)=:\cdots+\dt^2 X_{2,j,2}\,,
\]
\[
H_{3,j,k,l}=\dt Y_{3,j,k,l,1}+\dt^2\left(Y_{3,j,k,l,2} - X_{3,j,k,l,1} Z_1\right)=:\cdots+\dt^2 X_{3,j,k,l,2}\,,
\]
\[
H_{4,j,k}=\dt^{1/2}Y_{4,j,k,0}+\dt^{3/2}\left(Y_{4,j,k,1} - X_{4,j,k,0} Z_1\right)=:\cdots+\dt^{3/2}X_{4,j,k,1}\,,
\]
where 
\begin{equation}
Z_2=-\frac{i}2 X_{0,1} - \frac16 X_{0,0}^2 - \frac16 Q_1 + \frac{i}{24}  \{Q_0, X_{0,0}\} + \frac{1}{120} Q_0^2
\end{equation}
In addition,
\[
\begin{aligned}
H_0=&\dt^{1/2}\left(iY_{0,0}+i\frac{Q_0}{2}\right)+\dt^{3/2}\left(iY_{0,1}+\frac{i}{2}(Q_1+X^2_{0,0})-\frac{i}{24}Q^2_0+\frac{1}{6}\{Q_0,X_{0,0}\}\right)\\
&+\dt^{5/2}\left(iY_{0,2}+\frac{i}{2}\left(\{X_{0,0},X_{0,1}\}+Q_2\right)+\frac{1}{6}\left(X^3_{0,0}+\{Q_0,X_{0,1}\}+\{Q_1,X_{0,0}\}\right)\right.\\
    &-\frac{i}{24}\left(Q_0X^2_{0,0}+X_{0,0}Q_0X_{0,0}+X^2_{0,0}Q_0+\{Q_0,Q_1\}\right)\\
    &\left.-\frac{1}{120}\left(Q_0X_{0,0}Q_0+Q^2_0X_{0,0}+X_{0,0}Q^2_0\right)+\frac{i}{720}Q^3_0\right)
\end{aligned}\,,
\]
where
\begin{equation}
\begin{aligned}
Q_2=&\sum_{j=1}^{J} \left(X_{1,j,0}^\dag X_{1,j,2} + X_{1,j,2}^\dag X_{1,j,0}+X_{1,j,1}^\dag X_{1,j,1}+X_{2,j,1}^\dag X_{2,j,1}\right)+
\sum^J_{j,k,l=1}X_{3,j,k,l,1}^\dag X_{3,j,k,l,1}\\
&+ \sum^J_{j,k=1} \left(X_{4,j,k,1}^\dag X_{4,j,k,0}+X_{4,j,k,0}^\dag X_{4,j,k,1}\right)\,.
\end{aligned}
\end{equation}

\end{itemize}

\section{Proof of Lemma \ref{lem:R_alpha}}\label{sec:wt_R_alpha}
The purpose of \cref{lem:R_alpha} is to decompose the noise terms into uncorrelated random variables. 
According to \eqref{eqn:ploy_cov}, the noise terms in the It\^o-Taylor expansion \eqref{eqn:L} have the property that 
\begin{equation}\label{eqn:ploy_cov_2}
\mathbb{E}\left[R_{\mathrm{n},\alpha}R_{\mathrm{n},\alpha'}\right]=0,\quad \text{if}\ \alpha^+\neq(\alpha')^+\,.
\end{equation}
We define the set of multipositive indices $\Gamma^+_k$ as
\begin{equation}
\Gamma^+_k=\{\beta=(j_1,j_2,\cdots,j_{|\beta|})\in\{1,2,\cdots,J\}^{\otimes |\beta|}:|\beta|\leq k\}\,.
\end{equation}

Using the normalized noise, we rewrite $L[\ket{\psi}]$ in \eqref{eqn:L} as  
\[
L[\ket{\psi}]=\sum^k_{j=0}\frac{(\dt)^j}{j!}V^j_0\ket{\psi}+\sum_{\beta\in\Gamma^+_k}\sum_{\alpha^+=\beta}R_{\mathrm{n},\alpha} \left(\dt^{\frac{|\alpha|+l_{=0}(\alpha)}{2}}\textbf{V}_\alpha\ket{\psi}\right)\,.
\]
In light of \cref{eqn:ploy_cov_2}, to ensure zero correlation between random variables, it suffices to focus on the set $\{R_{\mathrm{n},\alpha}\}_{\alpha^+=\beta}$ for each $\beta\in\Gamma^+_k$.

In the remainder of the proof, we fix $\beta\in\Gamma^+_k$. To construct $\widetilde{R}$, we first fix an order of the noise terms $\left\{\widetilde{R}_{\mathrm{n},\alpha}\right\}_{\alpha^+=\beta}$ (the order can be arbitrary and does not affect the statement) and reformulate the sequence as $\{R_{\beta,i}\}_{i=1}^{I_\beta}$. Here $I_\beta$ denotes the cardinality of the set. Consequently, we rewrite the original summation $\left\{\widetilde{R}_{\mathrm{n},\alpha}\right\}_{\alpha^+=\beta}$, as
\[
\sum_{\alpha^+=\beta}R_{\mathrm{n},\alpha} \left(\dt^{\frac{|\alpha|+l_{=0}(\alpha)}{2}}\textbf{V}_\alpha\ket{\psi}\right)=:\sum_{i=1}^{I_\beta} R_{\beta,i} \left(\dt^{q_{\beta,i}}\textbf{V}_{\beta,i}\ket{\psi}\right)\,.
\]
We define $\textbf{V}_{\beta,i}=\textbf{V}_\alpha$ and $q_{\beta,i}=\frac{|\alpha|+l_{=0}(\alpha)}{2}$, where the index $i$ is assigned based on the specified ordering.

We define $\mathrm{Cov}_{\beta}$ as the covariance matrix of $\{R_{\beta,i}\}$. Because $\mathrm{Cov}_{\beta}$ is a positive semidefinite matrix, we can write $\mathrm{Cov}_{\beta}$ in eigendecomposition form $\mathrm{Cov}_{\beta}$  = $Q\Lambda Q^\top$, where $\Lambda$ is a diagonal matrix the entries of which are non-negative and $Q$ is an orthogonal matrix. We define
\[
\begin{bmatrix}
    \widetilde{R}_{\beta,1} \\
    \widetilde{R}_{\beta,2} \\
    \widetilde{R}_{\beta,3} \\
    \vdots \\
\end{bmatrix}= (\Lambda^+)^{-\frac{1}{2}}Q^\top\begin{bmatrix}
    R_{\beta,1} \\
    R_{\beta,2} \\
    R_{\beta,3} \\
    \vdots \\
\end{bmatrix}\,.
\]
where $\Lambda^{+}$ is a diagonal matrix such that
\[
(\Lambda^+)_{i,i}=\left\{\begin{aligned}
    &1,\quad \text{if}\ \Lambda_{i,i}=0\\
    &\Lambda_{i,i},\quad \text{if}\  \Lambda_{i,i}>0
\end{aligned}\right.\,.
\]
We have that $\left\{\widetilde{R}_{\beta,i}\right\}$ are not correlated, which means that $\mathbb{E}\left(\widetilde{R}_{\beta,i},\widetilde{R}_{\beta,j}\right)=0$ if $i\neq j$, and
\[
R_{\beta,i}=(\Lambda^+)_{i,i}\sum^i_{j=1} Q_{i,j}\widetilde{R}_{\beta,j}\,.
\]
where $\sum^i_{j=1}|Q_{i,j}|^2=\left(\mathrm{Cov}_{\beta}\right)_{i,i}=\mathbb{E}(R^2_{\beta,i})$. In addition, if $\widetilde{R}_{\beta,i}\neq 0$, then $\mathbb{E}(\widetilde{R}_{\beta,i}^2)=1$. This proves \eqref{eqn:R_alpha_lem}. 

\section{Proof of Lemma \ref{lem:H_existence}}\label{sec:H_existence}

Recall that $S_k+1=2^{a_t}$. To fulfill \eqref{eqn:H_lemma}, we need to construct a Hamiltonian
\begin{equation}\label{eq:Hmat}
\widetilde{H}= \begin{bmatrix}
    H_0 & H_1^\dag & H_2^\dag & \ldots & H_{S_k}^\dag \\
    H_1 & 0 & 0 & 0 & 0\\
    H_2 & 0 & 0 & 0 & 0\\
    \vdots  & 0 & 0 & 0 & 0\\
    H_{S_k} & 0 & 0 & 0 & 0
\end{bmatrix}
\end{equation}
that satisfies
\begin{equation}\label{eq:U2H}
    \bra{j}\exp(-i\sqrt{\dt}\widetilde{H}) \ket{0}=F_j+\mathcal{O}((\dt)^{k})\,.
\end{equation}
for $0\leq j\leq S_k$. 

Comparing \eqref{asymp-ham} with \eqref{asymp-kraus}, we reduce the power of $\dt$ by half because there is an extra $\sqrt{\dt}$ term in the Hamiltonian simulation~\eqref{asymp-ham}. We identify the blocks in \eqref{eq:Hmat} by asymptotically matching \eqref{asymp-kraus} and \eqref{asymp-ham}. For this purpose, we expand the matrix exponential in \eqref{asymp-kraus} using Taylor expansion:
\begin{equation}\label{eq: taylor}
  e^{-i \sqrt{\dt} \widetilde{H}} =  I - i\dt^{1/2}\widetilde{H} - \frac{\dt}{2} \widetilde{H}^2 + \frac{i\dt^{3/2}}{6}\widetilde{H}^3 + \frac{\dt^2}{24} \widetilde{H}^4 - \frac{i\dt^{5/2}}{120} \widetilde{H}^5 - \frac{\dt^3}{720} \widetilde{H}^6 + \cdots   
\end{equation}
To proceed, we first define,
\begin{equation}\label{eq: mat-Q}
    Q= \sum_{j=1}^{S_k} H_j^\dag H_j\,.
\end{equation}
This is the part of the operations that map the $\ket{0}$ ancilla to the $\ket{0}$ ancilla. Using \eqref{asymp-ham}, we can also expand $Q$ into an asymptotic form:
\begin{equation}\label{asymp-Q}
    Q= Q_0 + \dt Q_1 + \dt^2 Q_2 + \cdots
\end{equation}

We use asymptotic matching to obtain the form of $Q_l$ for all $l\le k$. Here $s_k$ is defined in~\eqref{asymp-kraus} and refers to the number of Kraus operators containing terms that scales as $\Delta t^{k+\frac12}$.
\begin{align}
        Q_0= &\sum_{j=1}^{s_k} X_{j,0}^\dag X_{j,0},\label{eqn:Q0} \\
        Q_1= & \sum_{j=1}^{s_k} \big( X_{j,0}^\dag X_{j,1} + X_{j,1}^\dag X_{j,0} \big) + \sum_{j=s_k+1}^{S_k} X_{j,0}^\dag X_{j,0}\,,\label{eqn:Q1}\\
        Q_2= & \sum_{j=1}^{s_k} \big( X_{j,0}^\dag X_{j,2} + X_{j,2}^\dag X_{j,0} +X_{j,1}^\dag X_{j,1}\big) + \sum_{j=s_k+1}^{S_k} \left(X_{j,1}^\dag X_{j,0}+X_{j,0}^\dag X_{j,1}\right)\,,\label{eqn:Q2}\\
        \cdots\notag\\
        Q_l= & \sum_{j=1}^{s_k} \sum^{l}_{p=0}\left( X_{j,p}^\dag X_{j,l-p}\right) + \sum_{j=s_k+1}^{S_k}\sum^{l-1}_{p=0} \left(X_{j,p}^\dag X_{j,l-1-p}\right)\,,\label{eqn:Qk}\\
        \cdots\notag
\end{align}

We determine the first term in each asymptotic expansion~\eqref{asymp-kraus}. We begin by matching the off-diagonal blocks in \eqref{eq:U2H}. Using \eqref{asymp-kraus}, \eqref{eq: taylor}, and \eqref{asymp-ham}, we have 
\begin{equation}\label{eqn:uniform_match_Y}
    Y_{j,0} + \dt Y_{j,1} + \dt^2 Y_{j,2}+\cdots = ( X_{j,0} + \dt X_{j,1} + \dt^2 X_{j,2} + \cdots ) (I + \dt Z_1 +  \dt^2 Z_2 + \cdots )
\end{equation}
for all $j>0$. Here, $\{Z_l\}^k_{l=1}$ are also operations that correspond to mapping the $\ket{0}$ ancilla to the $\ket{0}$ ancilla. They are defined as
\begin{equation}\label{eqn:Z}
\begin{aligned}
Z_1=& -\frac{i}{2} X_{0,0} - \frac16 Q_0\,, \\
Z_2=& -\frac{i}2 X_{0,1} - \frac16 X_{0,0}^2 - \frac16 Q_1 + \frac{i}{24}  \{Q_0, X_{0,0}\} + \frac{1}{120} Q_0^2\,.\\
\cdots\notag\\
Z_l=& -\frac{i}2 X_{0,l-1}+p_{z,l}(X_{0,0},X_{0,1},\cdots,X_{0,l-2},Q_0,Q_1,\cdots,Q_{l-1})\,,\\
\cdots\notag
\end{aligned} 
\end{equation}
where $p_{z,l}$ is a polynomial of degree $l$.

From the asymptotic analysis and the matching $\mathcal{O}(1)$ term in \eqref{eqn:uniform_match_Y}, we find the first coefficient in the off-diagonal blocks of the Hamiltonian matrix~\eqref{asymp-ham}:
\begin{align}
    X_{j,0} = Y_{j,0}, \quad j>0\,. \label{y2x1}
\end{align}

Next, we match the first block diagonal. By inserting the asymptotic expansion of $H_0$ and $Q$ into \eqref{eq: taylor}, we find
\begin{equation}\label{eqn:match_4_first}
\begin{aligned}
 I + \dt Y_{0,0}+\mathcal{O}(\dt^2)= I +\dt \left(-iX_{0,0}-\frac{Q_0}{2}\right)+\mathcal{O}(\dt^2)\,, 
\end{aligned}
\end{equation}
which leads to
\begin{equation}\label{eqn:X00}
X_{0,0}=iY_{0,0}+i\frac{Q_0}{2}\,.
\end{equation}
We now move on to the second term. Returning to \eqref{eqn:uniform_match_Y}, we can match the $\dt$ terms to obtain
\begin{equation}\label{eqn:second_order_term_match}
\begin{aligned}
X_{j,1} = & Y_{j,1} - X_{j,0} Z_1,\quad j>0\,.
\end{aligned}
\end{equation}
Additionally, equating the terms $\dt^2$ in the first block diagonal yields the second component of $H_0$ (for the sake of simplicity, we will not write down the asymptotic expansion):
\begin{equation}\label{eqn:X01}
X_{0,1}=iY_{0,1}+\frac{i}{2}(Q_1+X^2_{0,0})-\frac{i}{24}Q^2_0+\frac{1}{6}\{Q_0,X_{0,0}\}\,.
\end{equation}

To show that the above derivation process can always continue until we obtain the last term, we implement the induction argument. Assume that we have already matched $K$ terms and obtained
\begin{equation}
\begin{aligned}
    &X_{0,0},X_{0,1},\cdots,X_{0,K-1}\,,\\
    &X_{j,0},X_{0,1},\cdots,X_{j,K-1}\,,\quad j=1,2,\cdots,s_k\,,\\
    &X_{j,0},X_{0,1},\cdots,X_{j,K-1}\,,\quad j=s_k+1,\cdots,S_k\,.
\end{aligned}
\end{equation}
We can use the above terms, \eqref{eqn:Qk}, and \eqref{eqn:Z} to calculate
\begin{equation}
\begin{aligned}
    &Q_0,Q_1,\cdots,Q_{K-1}\,,\\
    &Z_0,Z_1,\cdots,Z_{K}\,.
\end{aligned}
\end{equation}
To continue, we first match the $\dt^{K}$ term in the off-diagonal blocks in \eqref{eq:U2H}. Similarly to \eqref{eqn:second_order_term_match}, we obtain
\begin{equation}\label{eqn:general_form_X1}
\begin{aligned}
&X_{j,K}=Y_{j,K}-\sum^K_{i=1}X_{j,K-i}Z_i, \quad j>0\,.
\end{aligned}
\end{equation}

Using $X_{j>0,k\leq K}$, we can construct $Q_K$ according to \eqref{eqn:Qk}. We then match the $\dt^{K+1}$ term in the first diagonal block in \eqref{eq:U2H}:
\begin{equation}\label{eqn:match_4}
\begin{aligned}
 &   I + \dt Y_{0,0} +\cdots + \dt^{K+1} Y_{0,k}+\mathcal{O}(\dt^{K+2})\\
= &I +\dt \left(-iX_{0,0}-\frac{Q_0}{2}\right) +\cdots+\dt^{K+1}\left(-iX_{0,K}+q_{x,K}(X_{0,0:K-1},Q_{0,0:K})\right)+\mathcal{O}(\dt^{K+2})\,.
\end{aligned}
\end{equation}
where $q_{x,K}$ is a polynomial of degree $K+1$. Thus, we obtain
\begin{equation}\label{eqn:general_form_X0}
X_{0,K}=iY_{0,K}-iq_{x,K}(X_{0,0:K-1},Q_{0,0:K})\,.
\end{equation}
This concludes the induction.

\begin{figure}
\centering
  \includegraphics[width=0.8\textwidth]{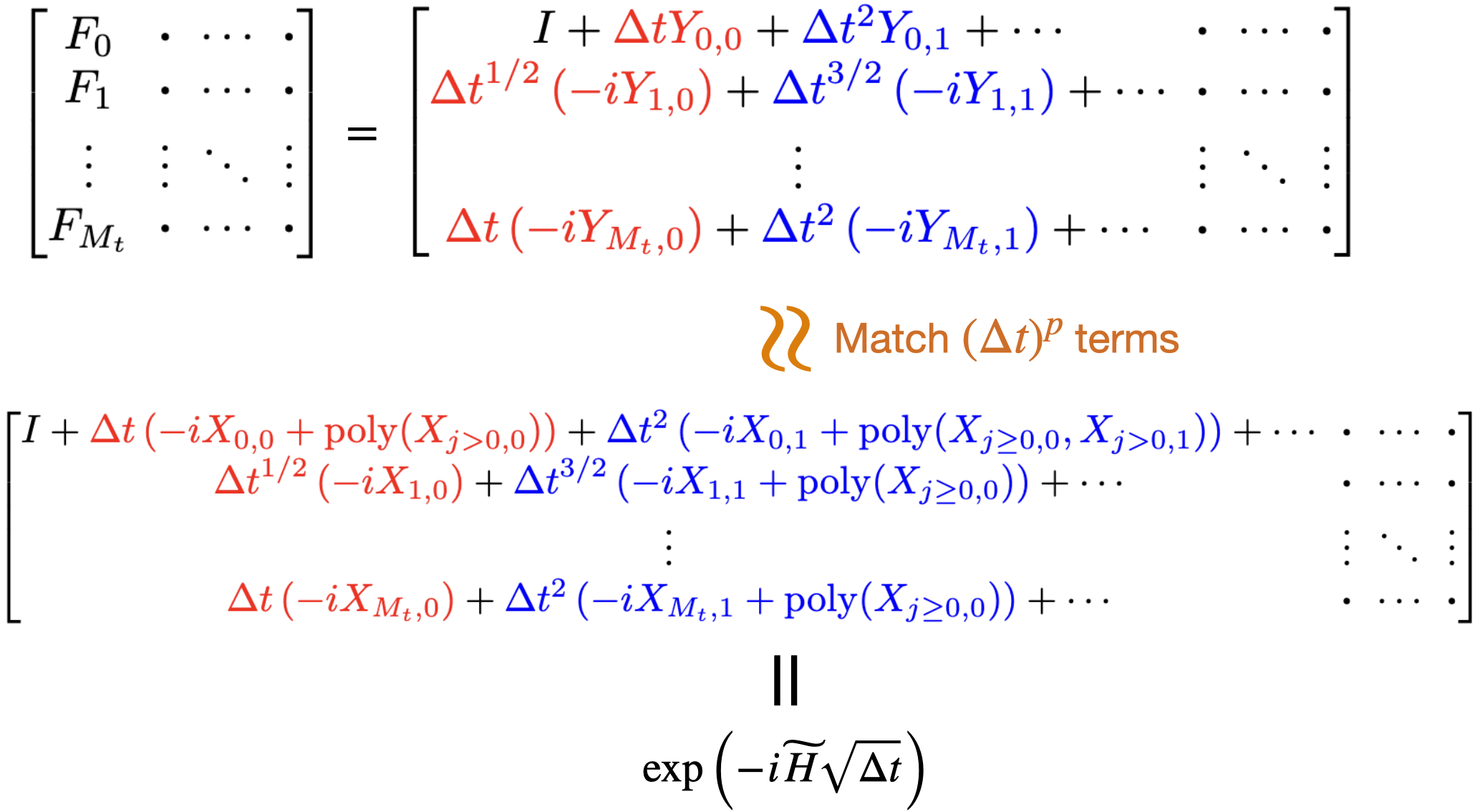}
   \caption{The generation of $\widetilde{H}$. We need to compare terms of the same order in the asymptotic expansion. Specifically, in each row of the two matrices, we need to match the terms with the same color. Here, $\mathrm{poly}([\cdot])$ means that the term can be written as a polynomial of elements in $[\cdot]$. } \label{fig:flowchart_match}
\end{figure}

In summary, to determine all the coefficients, we can follow the steps:
\begin{equation}\label{eqn:derivation_road}
\begin{aligned}
    \{Y_{j,0}\}_{j=1}^{s_k}&\underset{\eqref{y2x1}}{\to} \{X_{j,0}\}_{j=1}^{S_k}  \underset{\eqref{eqn:Q0}}{\to} Q_0 \underset{\eqref{eqn:X00}}{\to} X_{0,0}\underset{\eqref{eqn:Z}}{\to} Z_1  \\
    &\underset{\eqref{eqn:second_order_term_match}}{\to}\{ X_{j,1}\}_{j=1}^{S_k}\underset{\eqref{eqn:Q1}}{\to} Q_1 \underset{\eqref{eqn:X01}}{\to} X_{0,1}\underset{\eqref{eqn:Z}}{\to} Z_2 
 \\
&\cdots\\
&\underset{\eqref{eqn:general_form_X1}}{\to}\{ X_{j,k-1}\}_{j=1}^{s_k}
\underset{\eqref{eqn:Qk}}{\to} Q_{k-1}\underset{\eqref{eqn:general_form_X0}}{\to}X_{0,k-1}\,.
\end{aligned}
\end{equation}
For clarity, we provide a graph to show the generation of $\widetilde{H}$ in \cref{fig:flowchart_match}. Here, we note that in the last line, we only calculate $X_{j\leq s_k,k-1}$ because $Y_{j>s_k,k-1}=0$. 

By \eqref{eqn:F_formulation}, we find that each $Y_{j,q}$ is a polynomial of $H,V_j$ that satisfies $\|Y_{j,q}\|=\mathcal{O}(\|\mathcal{L}\|^{q+1/2}_{\mathrm{be}})$ for $1\leq j\leq s_k$ and $\|Y_{j,q}\|=\mathcal{O}(\|\mathcal{L}\|^{q+1}_{\mathrm{be}})$ otherwise. Inserting this into the preceding derivation, we find that each $X_{j,q}$ is a polynomial of $H$ and $V_j$ with the desired norm bound.

\section{\re{Proof of \texorpdfstring{\cref{lem:hermitian}}{Lg}}}\label{sec:lem:hermitian}
To show that $\widetilde{H}$ is a Hermitian matrix, we only need to prove that  $H_0$ is a Hermitian matrix.

We show this using the proof by contradiction. First, according to \eqref{eqn:trace_preserving}, we have
\begin{equation}
\left|\mathrm{Tr}\left(\sum^{S_k}_{j=0} F_j^\dagger\rho(0)F_j\right)-1\right|=\mathcal{O}(\dt^{k+1})
\end{equation}
for all $\rho(0)$. We define $\widetilde{U}=\exp(-i\widetilde{H}\sqrt{\dt})$. Then from \cref{lem:H_existence}, 
\begin{equation}
\left\|\Tr_A \left( \widetilde{U} \ketbra{0} \otimes \rho(0) \widetilde{U}^\dag \right)-\sum^{S_k}_{j=0} F_j^\dagger\rho(0)F_j\right\|_{1}=\mathcal{O}(\dt^{k+1}).
\end{equation}
This implies that
\begin{equation}\label{eqn:trace_one}
\left|\mathrm{Tr}\left( \widetilde{U} \ketbra{0} \otimes \rho(0) \widetilde{U}^\dag \right)-1\right|=\left|\mathrm{Tr}\left(\Tr_A \left( \widetilde{U} \ketbra{0} \otimes \rho(0) \widetilde{U}^\dag \right)\right)-1\right|=\mathcal{O}(\dt^{k+1})\,.
\end{equation}

If we assume that $H_0$ is non-Hermitian, it can be represented as:
\begin{equation}
H_{0}=D_{0}-iD_{1}(\Delta t)^p.
\end{equation}
In this expression, both $D_{0}$ and $D_{1}$ are Hermitian matrices. Additionally, $p$ satisfies $p \leq k - \frac{1}{2}$ and the norm of $D_{1}$ is of order one $1$: i.e., $\|D_{1}\|=\Omega(1)$.  Based on this representation, to construct $\widehat{H}$, one can extract the term $iD_{1}$ from $\widetilde{H}$ as
\begin{equation}
\widetilde{H}=\widehat{H}-i\ket{0}\bra{0}\otimes D_{1}(\dt)^{p},\quad \widetilde{U}=\exp(-i\sqrt{\dt}\widehat{H}-(\dt)^{p+1/2}\ket{0}\bra{0}\otimes D_{1})\,.
\end{equation}
Pick $\ket{\psi}$ such that $\|D_{1}\ket{\psi}\|=\Omega(1)$. We can apply Trotter splitting,
\begin{equation}
\left\|\widetilde{U}\ket{0}\otimes\ket{\psi}-\exp(-i\sqrt{\dt}\widehat{H})\exp(-(\dt)^{p+1/2}\ket{0}\bra{0}\otimes D_{1})\ket{0}\otimes\ket{\psi}\right\|=\mathcal{O}\left(\dt^{p+1}\right)\,,
\end{equation}
which leads to
\begin{equation}
\left|\|\widetilde{U}\ket{0}\otimes\ket{\psi}\|-\left\|\exp(-i\sqrt{\dt}\widehat{H})\exp(-(\dt)^{p+1/2}\ket{0}\bra{0}\otimes D_{1})\ket{0}\otimes\ket{\psi}\right\|\right|=\mathcal{O}\left(\dt^{p+1}\right)\,.
\end{equation}
Because $\|D_{1}\ket{\psi}\|=\Omega(1)$, there exists a constant $C$ independent of $\dt$ such that 
\begin{equation}
\left|\left\|\exp(-i\sqrt{\dt}\widehat{H})\exp(-(\dt)^{p+1/2}\ket{0}\bra{0}\otimes D_{1})\ket{0}\otimes\ket{\psi}\right\|-1\right|\geq C(\dt)^{p+1/2}\,.
\end{equation}
Combining the above two equalities, we obtain that there exists another constant $C'>0$ such that $\left|\|U\ket{0}\otimes\ket{\psi}\|-1\right|\geq C'(\dt)^{p+1/2}$. Since $p<k-1/2$, we conclude that
\begin{equation}
\left|\mathrm{Tr}\left( \widetilde{U} \ketbra{0} \otimes \ket{\psi}\bra{\psi} \widetilde{U}^\dag \right)-1\right|=\left|\|\widetilde{U}\ket{0}\otimes\ket{\psi}\|^2-1\right|=\Omega(\dt^{k})\,.
\end{equation}
which contradicts \eqref{eqn:trace_one}. This implies that $H_0$ must be a Hermitian matrix.

\section{\re{Block encoding of \texorpdfstring{$\widetilde{H}$}{TEXT}}}\label{sec:implement}
In this appendix, we describe a method to construct the block encoding of the dilated Hamiltonian $\widetilde{H}$. For simplicity, we assume access to the block encodings $\{U_j\}^{S_k}_{j=0}$ of $\{H_j\}^{S_k}_{j=0}$\footnote{Given that $H_j$ is a polynomial of $H$ and $V_j$, if we have the block encodings of $H$ and $V_j$, we can construct block encodings of $H_j$ using linear combination of unitaries.}. In particular, we have
\begin{equation}
\left(\bra{0_B}\otimes I_n\right)U_0\left(\ket{0_B}\otimes I_n\right)= \frac{H_0}{2D}\quad \text{and}\quad \left(\bra{0_B}\otimes I_n\right)U_j\left(\ket{0_B}\otimes I_n\right)= \frac{H_j}{D}\,,\quad \text{for $j>0$}\,,
\end{equation}
where $D=\Omega(\max_j\|H_j\|)$. At the end of this section, we will take a closer look at the derivation of $H_j$ and give an upper bound for $\|H_j\|$.

Without loss of generality, we also assume $S_k=2^P-1$ for some $P\in\mathbb{N}$. The construction of the block encoding of $\widetilde{H}$ can be divided into three steps:
\begin{itemize}
    \item Step 1: Construction of the block encoding of $\ketbra{0}\otimes \frac{H_0}{2}+\sum^{S_k}_{j=1}\ketbra{j}\otimes H^\dagger_j$. 

    We note that $\mathrm{U}=\sum^{S_k}_{j=0}\ketbra{j}\otimes U^\dagger_j$ provides a block encoding of $\ketbra{0}\otimes \frac{H_0}{2}+\sum^{S_k}_{j=1}\ketbra{j}\otimes H^\dagger_j$ by the following equation:
    \begin{equation}
    \left(I_{A}\otimes\bra{0_B}\otimes I_n\right)\sum^{S_k}_{j=0}\ketbra{j}\otimes U^\dagger_j\left(I_{A}\otimes\ket{0_B}\otimes I_n\right)= \frac{1}{D}\left(\ketbra{0}\otimes \frac{H_0}{2}+\sum^{S_k}_{j=1}\ketbra{j}\otimes H^\dagger_j\right)\,,
    \end{equation}
    where $I_A$ is the identity map that acts on the $P$ ancilla qubits. In the worst case, this selected oracle $\mathrm{U}$ can be  constructed using $S_k+1$ controlled logic gates. We give an example with $S_k=3$ in \cref{fig:circuit_1}:
    \begin{figure}[H]
    \centering
    \begin{quantikz}[column sep=0.3cm]
    \qw &\octrl{1}&\octrl{1}&\ctrl{1}&\ctrl{1}&\qw\\
    \qw &\octrl{1}&\ctrl{1}&\octrl{1}&\ctrl{1}&\qw \\
   \qw&\gate[2]{U_0}&\gate[2]{U_1}&\gate[2]{U_2}&\gate[2]{U_3}&\qw\\
    \qw&&&&&\qw
    \end{quantikz}
    \caption{Quantum circuit for directly implementing $\mathrm{U}=\sum^{3}_{j=0}\ketbra{j}\otimes U^\dagger_j$.}
    \label{fig:circuit_1}
    \end{figure}

    \item Construction of the block encoding of $\ketbra{0}\otimes \frac{H_0}{2}+\sum^{S_k}_{j=1}\ketbra{0}{j}\otimes H^\dagger_j$.

    To construct the block encoding of $\ketbra{0}\otimes \frac{H_0}{2}+\sum^{S_k}_{j=1}\ketbra{0}{j}\otimes H^\dagger_j$, we apply a block encoding of $\sum^{S_k}_{j=0}\ketbra{0}{j}\otimes I_B\otimes I_n$ to $\mathrm{U}$. In particular, add $P$ ancilla qubits and define the operator
    \begin{equation}
    \mathrm{W}=\left(\mathrm{H}^{\otimes P}\otimes I_A\otimes I_B\otimes I_n\right)\mathrm{SWAP}_{A}\otimes I_B\otimes I_n\,,
    \end{equation}    
    where $\mathrm{SWAP}_{A}\ket{0_A}\ket{b_A}=\ket{b_A}\ket{0_A}$ and $\mathrm{H}$ is the Hadamard gate that is used to recover $\frac{1}{\sqrt{S_k+1}}\ket{0_A}+\ket{\perp}$. The unitary gate $\mathrm{W}$ can be implemented using $P$ Hadamard gates and $3^P$ controlled logic gates. We draw the circuit in \cref{fig:circuit_2}.
    
    We note that $\mathrm{W}$ satisfies our requirement, which means that \begin{equation}
    \mathrm{W}
    \left(I_A\otimes \mathrm{U}\right)=\frac{1}{\sqrt{S_k+1}}\ketbra{0_A}\otimes 
    \left(\sum^{S_k}_{j=0}\ketbra{0}{j}\otimes U^\dagger_j\right)+\ketbra{\perp}\,.
    \end{equation}
    Furthermore, plugging the formula of $\mathrm{U}$, we obtain
    \begin{equation}
    \begin{aligned}
            &\left(\bra{0_A}\otimes I_{A}\otimes\bra{0_B}\otimes I_n\right)\mathrm{W}\left(I_A\otimes \mathrm{U}\right)\left(\ket{0_A}\otimes I_{A}\otimes\ket{0_B}\otimes I_n\right)\\
            =&\frac{1}{\sqrt{S_k+1}D}\left(\ketbra{0}\otimes \frac{H_0}{2}+\sum^{S_k}_{j=1}\ketbra{0}{j}\otimes H^\dagger_j\right)\,.
    \end{aligned}
    \end{equation}
    Thus, $\mathrm{W}\left(I_{A}\otimes \mathrm{U}\right)$ is the block encoding of $\ketbra{0}\otimes \frac{H_0}{2}+\sum^{S_k}_{j=1}\ketbra{0}{j}\otimes H^\dagger_j$.

    \begin{figure}[H]
    \centering
    \begin{quantikz}[column sep=0.3cm]
    \qw &\targ{}& \ctrl{1}&\targ{}&\gate{\mathrm{H}^{\otimes P}}&\qw\\
    \qw &\ctrl{-1}&\targ{}&\ctrl{-1}&\qw&\qw
    \end{quantikz}
    \caption{Quantum circuit for $\mathrm{W}=\left(\mathrm{H}^{\otimes P}\otimes I_A\right)\mathrm{SWAP}_{A}$.}
    \label{fig:circuit_2}
    \end{figure}

    \item Construction of the block encoding of $\ketbra{0}\otimes H_0+\sum^{S_k}_{j=1}\left(\ketbra{j}{0}\otimes H_j+\ketbra{0}{j}\otimes H^\dagger_j\right)$.

    This step can be completed by an LCU circuit, as drawn in \cref{fig:circuit_3}. More specifically, the circuit implements the block encoding of $\left(\mathrm{W}\left(I_{A}\otimes \mathrm{U}\right)\right)+\left(\mathrm{W}\left(I_{A}\otimes \mathrm{U}\right)\right)^\dagger$. 
    
    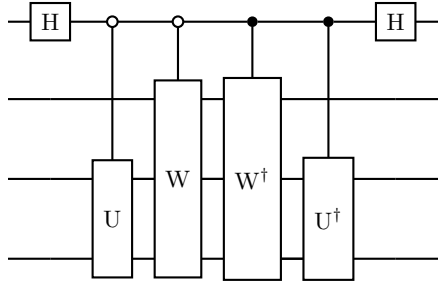
\begin{figure}[H]
    \centering
        \begin{quantikz}[column sep=0.3cm]
    \qw &\gate{\mathrm{H}}&\octrl{2}&\octrl{1}&\ctrl{1}&\ctrl{2}&\gate{\mathrm{H}}&\qw\\
    \qw &\qw&\qw&\gate[3]{\mathrm{W}}&\gate[3]{\mathrm{W}^\dagger}&\qw&\qw&\qw \\
   \qw&\qw&\gate[2]{\mathrm{U}}&\qw&\qw&\gate[2]{\mathrm{U}^\dagger}&\qw&\qw\\
    \qw&\qw&&&&\qw&\qw&\qw
    \end{quantikz}
    \caption{Quantum circuit for the block encoding of  $\widetilde{H}$.}
    \label{fig:circuit_3}
    \end{figure}

\end{itemize}

In summary, define the operator generated by \cref{fig:circuit_3} as $\mathrm{Q}$. According to the above derivation, $\mathrm{Q}$ is a block encoding of $\widetilde{H}$, meaning 
\begin{equation}
    \left(\bra{0}\otimes\bra{0_A}\otimes I_{A}\otimes\bra{0_B}\otimes I_n\right) \mathrm{Q}\left(\ket{0}\otimes\ket{0_A}\otimes I_{A}\otimes\ket{0_B}\otimes I_n\right)=\frac{1}{\sqrt{2(S_k+1)}D}\widetilde{H}.
\end{equation}

The success probability of the block encoding is inversely proportional to $(S_k+1)D^2$. The next step is to determine an upper bound for $D$, which is equivalent to finding the maximum value of $\|H_j\|$.

We first consider the norm of the asymptotic expansion term $Y_{j,q}$ defined in \eqref{asymp-kraus}. We upper bound $\|Y_{j,q}\|$ in the following lemma:
\begin{lem} Fix $k\geq 1$. Given $0\leq j\leq S_k$ and $0\leq q\leq k$,
\begin{equation}\label{eqn:Y_j_q_bound}
    \|Y_{j,q}\|=\left\{\begin{aligned}
    &(4\|\mathcal{L}\|_{\mathrm{be}})^{q+1/2},\quad \forall 1\leq j\leq s_k,\\
    &(4\|\mathcal{L}\|_{\mathrm{be}})^{q+1},\quad \text{otherwise.}
\end{aligned}\right.
\end{equation}
\end{lem}
\begin{proof}
Recall the construction of $F$ in \eqref{eqn:F_formulation}. For any $\alpha\in \Gamma_{k/0}$ and $\dt^p$ terms, where $2p\in\mathbb{N}$ and $0\leq 2p\leq 2k+1$, the coefficient has the bound
\begin{equation}\label{eqn:coefficient_H_bound}
\sum_{\frac{|\alpha'|+l_{=0}(\alpha')}{2}=p,(\alpha')^+=\alpha^+}\left|c_{\alpha',\alpha}\right|\leq \sum_{\frac{|\alpha'|+l_{=0}(\alpha')}{2}=p,(\alpha')^+=\alpha^+}C^{1/2}_{\alpha',\alpha'}\leq \sum_{\frac{|\alpha'|+l_{=0}(\alpha')}{2}=p,(\alpha')^+=\alpha^+}1\leq 4^p\,.
\end{equation}
Here, we use $|c_{\alpha',\alpha}|^2\leq \mathbb{E}(R^2_{\mathrm{n},\alpha'})=C_{\alpha',\alpha'}\leq 1$ in the first and second inequalities. Combining this and $\left\|\textbf{V}_{\alpha'}\right\|_2\leq \|\mathcal{L}\|^{|\alpha'|}_{be}$, we prove \eqref{eqn:Y_j_q_bound}.
\end{proof}

Next, we recall the formula of $H_j$ in \cref{lem:H_existence} \eqref{asymp-ham}. To bound $\|H_j\|$, we first give the bound for $\|X_{j,q}\|$ in the following lemma:
\begin{lem}\label{lem:X_j_q_bound} Fix $k\geq 1$. Assume $0\leq j\leq S_k$ and $0\leq q\leq k$, then
\begin{equation}\label{eqn:bound_X_j_q}
    \left\|X_{j,q}\right\|\leq \left\{
    \begin{aligned}
    &\left(4(q+1)!(J+1)^{4k(q+1/2)}\|\mathcal{L}\|_{\mathrm{be}}\right)^{q+1/2},\quad 1\leq j\leq s_k\,,\\
    &\left(4(q+1)!(J+1)^{4k(q+1)}\|\mathcal{L}\|_{\mathrm{be}}\right)^{q+1},\quad \text{otherwise}\,.
    \end{aligned}\right.
\end{equation}
\end{lem}

\begin{proof} According to the matching equation~\eqref{eq:U2H} and the asymptotic form~\eqref{asymp-kraus}, \eqref{asymp-ham}, in $\widetilde{H}$ we should have terms $X_{j,q}\Delta t^{q}$ for $1\leq j\leq s_k$ and $X_{j,q}\Delta t^{q+{1/2}}$ for other $j$. Furthmore, by matching the power of $\Delta t$ on both sides of~\eqref{eq:U2H}, we can rewrite \eqref{eqn:general_form_X1} and \eqref{eqn:general_form_X0} as
\begin{equation}\label{eqn:iteration_new}
X_{j,q}=\left\{
\begin{aligned}
&iY_{0,q}+\sum_{\xi\in\Xi^j_q,\gamma\in \Pi^j_{q}}c_{\xi,\gamma;j,q} X_{0,\gamma_1}X^\dagger_{\xi_1,\gamma_2}X_{\xi_2,\gamma_3}\cdots,\quad j=0\\
&Y_{j,q}+\sum_{\xi\in\Xi^j_q,\gamma\in \Pi^j_{q}}c_{\xi,\gamma;j,q} X_{j,\gamma_1}X^\dagger_{\xi_1,\gamma_2}X_{\xi_2,\gamma_3}\cdots,\quad j>0
\end{aligned}\right.\,.
\end{equation}
Here $|c_{\xi,\gamma;j,q}|\leq 1$ and $c_{\xi,\gamma;j,q}=0$ if $|\xi|\neq |\gamma|-1$, 
\[
\Xi^j_q=\left\{\xi=(\xi_1,\cdots,\xi_{|\xi|})\in\{0,1,2,\cdots,S_k\}^{\otimes |\xi|}:|\xi|\leq 2q-1\right\}
\]
and 
\[
\Pi^j_{q}=\left\{
\begin{aligned}
&\left\{\gamma=(\gamma_1,\cdots,\gamma_{|\gamma|})\in\{0,1,2,\cdots,q-1\}^{\otimes |\gamma|}:\sum^{|\gamma|}_{i=1}(\gamma_i+1/2)\leq q+1/2\right\},\quad 1\leq j\leq s_k\\
&\left\{\gamma=(\gamma_1,\cdots,\gamma_{|\gamma|})\in\{0,1,2,\cdots,q-1\}^{\otimes |\gamma|}:\sum^{|\gamma|}_{i=1}(\gamma_i+1/2)\leq q+1\right\},\quad \text{otherwise}
\end{aligned}\,.
\right.
\]
for $q>0$ and $\Pi^j_0=\emptyset$. We note that to match the power of $\Delta t$, $c_{\xi,\gamma;j,q}=0$ whenever the power of $\Delta t$ corresponds to $X_{j,\gamma_1}X^\dagger_{\xi_1,\gamma_2}X_{\xi_2,\gamma_3}\dots$ exceeds $q+1-\frac{\textbf{1}_{1\leq j\leq s_k}+|\gamma|}{2}$.

We prove \eqref{eqn:bound_X_j_q} by induction. From \eqref{eqn:Y_j_q_bound}, we obtain that, when $q=0$, \eqref{eqn:bound_X_j_q} is true. 

Assume that \eqref{eqn:bound_X_j_q} is true for $q\leq Q-1$:
\begin{itemize}
\item Fix $1\leq j\leq s_k$. Using \eqref{eqn:iteration_new},
\[
\begin{aligned}
    \left\|X_{j,Q}\right\|&\leq \left\|Y_{j,Q}\right\|+\sum_{\gamma\in \Pi_{Q}}c_{\xi,\gamma;j,Q} \left((J+1)^k\right)^{|\gamma|-1}\left(4Q!(J+1)^{4kQ}\|\mathcal{L}\|_{\mathrm{be}}\right)^{Q+1/2}\\
    &<(4\|\mathcal{L}\|_{\mathrm{be}})^{Q+1/2}+\left((J+1)^{4k(Q+1/2)}\right)^{Q+1/2}\left(4Q!\|\mathcal{L}\|_{\mathrm{be}}\right)^{Q+1/2}\sum_{\gamma\in\Pi_Q}1\\
        & \leq (4\|\mathcal{L}\|_{\mathrm{be}})^{Q+1/2}+\left((J+1)^{4k(Q+1/2)}\right)^{Q+1/2}\left(4Q!Q\|\mathcal{L}\|_{\mathrm{be}}\right)^{Q+1/2}\\
        &\leq \left(4(Q+1)!(J+1)^{4k(Q+1/2)}\|\mathcal{L}\|_{\mathrm{be}}\right)^{Q+1/2}
\end{aligned}\,.
\]
Here, we use the induction bound and $|\left\{\xi|\xi\in\Xi_q,|\xi|=|\gamma|-1\right\}|=((J+1)^k)^{|\gamma|-1}$ in the first inequality. Furthermore, the power of $Q+\frac{1}{2}$ comes from the fact that the power of $\Delta t$ corresponds to $X_{j,\gamma_1}X^\dagger_{\xi_1,\gamma_2}X_{\xi_2,\gamma_3}\dots$ cannot exceed $Q+\frac{1}{2}-\frac{|\gamma|}{2}$, which implies that the power of $\left(4Q!(J+1)^{2k}\|\mathcal{L}\|_{\mathrm{be}}\right)$ cannot exceed $Q+\frac{1}{2}$. In the second inequality, we use \eqref{eqn:Y_j_q_bound}, $|c_\gamma|<1$, and $|\gamma|\leq 2Q$. In the third inequality, we use
$|\Pi_Q|\leq Q^Q$. 

\item Fix $j=0$ or $j>s_k$. Similar to before, using \eqref{eqn:iteration_new},
\[
\begin{aligned}
    \left\|X_{j,Q}\right\|&\leq \left\|Y_{j,Q}\right\|+\sum_{\gamma\in \Pi_{Q}}c_{\xi,\gamma;j,Q} \left((J+1)^k\right)^{|\gamma|-1}\left(4Q!(J+1)^{4kQ}\|\mathcal{L}\|_{\mathrm{be}}\right)^{Q+1}\\
    &<(4\|\mathcal{L}\|_{\mathrm{be}})^{Q+1}+\left((J+1)^{4k(Q+1)}\right)^{Q+1}\left(4Q!\|\mathcal{L}\|_{\mathrm{be}}\right)^{Q+1}\sum_{\gamma\in\Pi_Q}1\\
        & \leq (4\|\mathcal{L}\|_{\mathrm{be}})^{Q+1}+\left((J+1)^{4k(Q+1)}\right)^{Q+1}\left(4Q!Q\|\mathcal{L}\|_{\mathrm{be}}\right)^{Q+1}\\
        &\leq \left(4(Q+1)!(J+1)^{4k(Q+1)}\|\mathcal{L}\|_{\mathrm{be}}\right)^{Q+1}
\end{aligned}\,.
\]
\end{itemize}
The above two inequalities conclude the induction and prove \eqref{eqn:bound_X_j_q}.
\end{proof}

Finally, using \cref{lem:X_j_q_bound} \eqref{eqn:bound_X_j_q}, it is straightforward to obtain
\[
\left\|H_j\right\|\leq \left\{\begin{aligned}
&\sum^{k-1}_{q=0} \left(4(q+1)!(J+1)^{4k(q+1/2)}\|\mathcal{L}\|_{\mathrm{be}}\right)^{q+1/2}\dt^{q},\quad 1\leq j\leq s_k\,,\\    
&\sum^{k-1}_{q=0} \left(4(q+1)!(J+1)^{4k(q+1)}\|\mathcal{L}\|_{\mathrm{be}}\right)^{q+1}\dt^{q+1/2},\quad \text{otherwise}
\end{aligned}\right.
\]
for all $0\leq j\leq S_k$. Choosing $\dt=\mathcal{O}\left(\left(k^{k
+2}(J+1)^{4(k+1)^2}\|\mathcal{L}\|_{\mathrm{be}}\right)^{-1}\right)$, the above equation suggests that
\begin{equation}\label{eqn:alpha_bound}
\begin{aligned}
D=&\mathcal{O}\left((J+1)^{2k}\|\mathcal{L}\|_{\mathrm{be}}^{1/2}\sum^{k-1}_{q=1} \left(\left(4(q+1)!(J+1)^{4k(q+1)}\right)^{1+1/q}\|\mathcal{L}\|_{\mathrm{be}}\right)^{q}\dt^{q}\right)\\
=&\mathcal{O}\left((J+1)^{2k}\|\mathcal{L}\|_{\mathrm{be}}^{1/2}\right)
\end{aligned}\,.
\end{equation}
Recall $S_k=\mathcal{O}((J+1)^{k+1})$ from \cref{thm:main}. We obtain the bound for the subnormalization factor of the block encoding: 
\begin{equation}\label{eqn:subnormalization_factor}
\sqrt{2(S_k+1)}D=\mathcal{O}\left((J+1)^{(5k+1)/2}\|\mathcal{L}\|^{1/2}_{\rm be}\right)\,.
\end{equation}
It should be noted that the bound presented in \cref{eqn:alpha_bound} significantly overstates the size of $\norm{H_j}$. In the proof of \cref{lem:X_j_q_bound}, we upper bound the small coefficient $c$ by $1$ and do not take into account possible cancelation of terms in the summation. In practice, considerable cancelation occurs within asymptotic expansion and matching, resulting in a substantially reduced $\norm{H_j}$. 

A concrete illustration of this idea can be seen from the first order to the third order provided in \cref{sec:all_scheme}. Let us consider $k=2$ for example, where the dilated Hamiltonian $\widetilde{H}$ takes the form of \cref{eqn:second_order_appendix'}. We have  
\[
\left\|H_0\right\|_2\leq\sqrt{\Delta t} \|\mathcal{L}\|_{be}+\frac{\Delta t^{3/2}}{12}\|\mathcal{L}\|_{be}^2\,,\quad \left\|H_j\right\|_2\leq \|\mathcal{L}\|^{1/2}_{be}+\frac{5\Delta t}{6}\|\mathcal{L}\|^{3/2}_{be},\quad 1\leq j\leq J\,.
\]
\[
\left\|H_{j+J}\right\|_2\leq \frac{\Delta t}{\sqrt{12}}\|\mathcal{L}\|^{3/2}_{be},\quad 1\leq j\leq J\,.
\]
\[
\left\|H_{2J+(l-1)J^2+(k-1)J+j}\right\|_2\leq \frac{\Delta t}{\sqrt{6}}\|\mathcal{L}\|^{3/2}_{be},\quad 1\leq j,k,l\leq J\,.
\]
\[
\left\|H_{J^3+2J+(k-1)J+j}\right\|_2\leq \sqrt{\frac{\Delta t}{2}}\|\mathcal{L}\|_{be},\quad 1\leq j,k\leq J\,.
\]
We note that $S_2=J^3+J^2+J$ and $S_2+1\leq (J+1)^3$. Then, we could choose $\Delta t=\frac{5}{6}\|\mathcal{L}_{be}\|^{-1}$ and $D=2\|\mathcal{L}\|^{1/2}_{\rm be}$. This gives us the subnormalization factor:
\[
\sqrt{2(S_k+1)}D=2\sqrt{2}(J+1)^{3/2}\|\mathcal{L}\|^{1/2}_{\rm be}\,.
\] 
We observe that the choice $\Delta t$ is independent of $J$, and the power of $J$ in the subnormalization factor is significantly smaller than that of \eqref{eqn:subnormalization_factor}.

\end{document}